\documentclass[preprint2]{aastex6}

\usepackage{subfigure}
\usepackage{amsmath}
\usepackage{mathtools}
\usepackage[all]{hypcap}
\begin{document}


\title{Rotation-supported neutrino-driven supernova explosions in three dimensions\\ and the critical luminosity condition}


\author{Alexander Summa\altaffilmark{1}, Hans-Thomas Janka\altaffilmark{1}, Tobias Melson\altaffilmark{1}, and Andreas Marek\altaffilmark{2}}
\altaffiltext{1}{Max-Planck-Institut f\"ur Astrophysik, Karl-Schwarzschild-Str.~1, D-85748 Garching, Germany, \\email: thj@mpa-garching.mpg.de}
\altaffiltext{2}{Max Planck Computing and Data Facility (MPCDF), Gie{\ss}enbachstr.~2, D-85748 Garching, Germany}


%
%





\begin{abstract}
We present the first self-consistent, three-dimensional (3D) core-collapse 
supernova simulations performed with the \textsc{Prometheus-Vertex} code 
for a rotating progenitor star. Besides using the angular momentum of the
15\,M$_\odot$ model as obtained in the stellar evolution calculation with
an angular frequency of $\sim$\,10$^{-3}$\,rad\,s$^{-1}$ (spin period of more
than 6000\,s) at the Si/Si-O interface, we also computed
2D and 3D cases with no rotation and with a $\sim$\,300 times shorter rotation 
period and different angular resolutions. In 2D, only the
nonrotating and slowly rotating models explode, while rapid rotation prevents
an explosion within 500\,ms after bounce because of lower radiated neutrino
luminosities and mean energies and thus reduced neutrino heating.
In contrast, only the fast rotating model develops an explosion in 3D when
the Si/Si-O interface collapses through the shock. The explosion becomes
possible by the support of a powerful SASI spiral mode, which compensates
for the reduced neutrino heating and pushes strong shock expansion in the 
equatorial plane. Fast rotation in 3D leads to a ``two-dimensionalization''
of the turbulent energy spectrum (yielding roughly a $-$3 instead of a $-$5/3 
power-law slope at intermediate wavelengths) with enhanced 
kinetic energy on the largest spatial scales.
We also introduce a generalization of the ``universal critical luminosity
condition'' of \citet{Summa2016} to account for the effects of rotation, and
demonstrate its viability for a set of more than 40 core-collapse simulations
including 9 and 20\,M$_\odot$ progenitors as well as black-hole forming cases
of 40 and 75\,M$_\odot$ stars to be discussed in forthcoming papers.
\end{abstract}

\keywords{supernovae: general --- hydrodynamics --- instabilities --- neutrinos}

\section{Introduction} \label{sec:intro}

Although recent three-dimensional (3D) simulations of core-collapse supernovae (CCSN) were able to 
yield successful explosions for a variety of progenitor stars 
and showed the viability of the neutrino-driven explosion mechanism in principle
\citep[see, e.g.,][]{Takiwaki2012,Takiwaki2014,Lentz2015,Melson2015,Melson2015a,Mueller2016,Roberts2016},
3D models seem to be less prone to explosion than their corresponding axisymmetric 
(2D) counterparts \citep[for an extensive review see][]{Janka2016}. 
Due to the extreme computational demands of 3D simulations including
energy-dependent neutrino transport treatments, the numerical resolution in current 3D
models is still limited. While the influence of this deficit on the explosion physics
remains to be investigated in fully self-consistent supernova simulations, 
the enhanced reluctance to explosion in 3D may also point
towards important physical ingredients that have not yet been considered in the latest
3D simulations.

In order to get robust and sufficiently energetic neutrino-powered explosions, several 
solutions providing additional support to the neutrino-driven mechanism have been proposed
up to now. Even though many of these ideas have already been investigated in the context
of 1D and 2D simulations, final conclusions heavily rely on simulations without imposed
symmetry constraints because of the three-dimensional nature of the expected effects. 
The era of sophisticated 3D simulations has just begun and first steps towards the inclusion of 
additional physics have been made recently. These comprise the consideration of 
progenitor properties such as pre-collapse perturbations \citep{Couch2013a,Couch2015a,Mueller2015,Mueller2016a,Mueller2016,Mueller2017}, 
rotation \citep{Kuroda2014,Nakamura2014,Takiwaki2016}, and magnetic fields \citep{Winteler2012,Moesta2014}, 
as well as the exploration of 
changes in the microphysics, which could enhance the neutrino heating behind the shock
\citep{Melson2015a,Bollig2017}.

In the present work we focus on the possible support of neutrino-driven supernova explosions 
by rotation. The effects of rotation are of great interest in the context of CCSNe, 
because the angular momentum of the stellar core 
can influence both the explosion physics and the birth properties of the neutron star \citep[e.g.][]{Ott2006}.
While rapid rotation is a key prerequisite for explosion scenarios such as the magnetorotational
mechanism \citep{Akiyama2003,Dessart2007,Winteler2012,Moesta2014,Moesta2015,Obergaulinger2017} or models developing the low-$T/W$ spiral 
instability \citep{Takiwaki2016}, large rotation rates
are generally not expected for progenitors of typical type IIP CCSNe \citep{Heger2005}. 
Asteroseismic measurements of evolved
low-mass stars point towards an efficient spin-down mechanism of the core \citep{Cantiello2014}. Models
including the angular momentum transport by magnetic fields predict pre-collapse angular momenta with peak values of the 
order of $10^{15}\,\mathrm{cm}^2\,\mathrm{s}^{-1}$ at a mass coordinate of $1.5\,\mathrm{M}_\odot$
\citep{Heger2005}. Furthermore, no efficient mechanism is known which could slow down
a new-born \mbox{(sub-)millisecond-period} neutron star resulting from a rapidly rotating iron core 
to the typical observed periods of tens to hundreds of milliseconds of young pulsars \citep{Ott2006}.
Simplified simulations point towards a subtle interplay between rotation and the growth
of the spiral mode of the standing accretion-shock instability (SASI) even for moderate rotation 
rates \citep{Blondin2007,Iwakami2009,Blondin2017,Kazeroni2017}, calling for thorough investigations with sophisticated
 explosion models taking into account full neutrino physics.

For the first time, we present the results of rotating, self-consistent 3D
simulations with the \textsc{Prometheus-Vertex} code applying the most complete set of neutrino 
interactions currently available. In our study, we consider a $15\,\mathrm{M}_\odot$ progenitor from the 
stellar evolution calculations of \citet{Heger2005}, which include rotation and angular momentum transport 
by magnetic fields. In addition to the rotation profile inferred from \citet{Heger2005}, we perform
simulations for the same progenitor, but with enhanced angular velocities as previously applied in \citet{Buras2006b} 
and \citet{Marek2009a}. In total, our study consists of five 3D and six 2D models with 
different resolutions and rotation rates. Since many 2D studies concerning the influence of rotation 
on the explosion physics and the final neutron star spin have already been conducted up to now 
\citep[see, e.g.,][]{Kotake2006,Ott2006}, we only provide a brief comparison of the axisymmetric 
models to their 3D counterparts and center our discussion of the results on
the two best resolved 3D simulations.

In multidimensional simulations, the beginning of runaway shock expansion seems to be characterized by a huge
model-to-model variance in the diagnostic parameters such as neutrino luminosity, neutrino heating rate and efficiency, 
shock radius, mass-accretion rate, turbulent kinetic and total energy in the gain layer, and so on.
This naturally raises the question if any combination of the relevant 
parameters reliably signals the onset of shock expansion.
Extending the critical luminosity condition introduced by \citet{Burrows1993} from spherical symmetry to
the multidimensional case, \citet{Summa2016} showed that the consideration of turbulent mass motions in 
the gain layer through an isotropic pressure contribution \citep[cf.][]{Mueller2015} leads to a 
generalized criterion which captures the conditions needed for shock runaway  in a large set of 2D models remarkably well.

A reduction of the critical luminosity of shock runaway due to rotation and SASI spiral modes has been 
demonstrated with 3D simulations by \citet{Iwakami2014} and \citet{Nakamura2014}. However,
both studies use a simple light-bulb treatment for the neutrino emission, which is unable to capture
the feedback effects of rotation on the neutrino transport. In contrast, our models are fully self-consistent in the sense that 
they evolve the neutrino transport coupled with the hydrodynamics
of the stellar plasma. Moreover, a subset of our 2D and 3D simulations 
are started from the angular momentum profile provided by a stellar
evolution calculation up to the onset of iron-core collapse 
performed by \citet{Heger2005}.

The coupling of neutrino transport and hydrodynamics allows us to 
take into account changes of the neutrino emission and of the
associated postshock neutrino heating that are caused by 
rotational effects. For example, rapid rotation has consequences for
the structure of the neutron star (e.g., it leads to centrifugal deformation),
for the convective activity inside of the neutron star \citep[because
steep angular momentum gradients can suppress convection; see][]
{Buras2006b,Dessart2006}, and for the mass flow
through the postshock layer onto the nascent neutron star (since, e.g.,
SASI sloshing and spiral modes affect the dynamics of the postshock
region and trigger shock expansion). All of these phenomena lead to
modifications of the radiated neutrino luminosities and spectra and
to corresponding directional variations, which are not accounted
for in simple light-bulb treatments. Employing a pre-collapse
angular momentum distribution from a stellar evolution model defines
an important reference point for the influence of rotation on
the explosion in ``realistic'' stellar cores. Our results will 
demonstrate that all of these aspects, which reach beyond previous
studies using neutrino light bulbs and parametrized initial
rotation profiles, are relevant for assessing the importance of
rotation in the neutrino-heating mechanism for driving SN explosions.

Similar to the correction factor accounting for turbulent pressure and thermal dissipation, we derive an additional term 
representing the effects of rotation \citep[see also][]{Janka2016}. This allows us to include our set of rotation models
to the representation of the generalized critical condition introduced by \citet{Summa2016}. In order to
validate this criterion for an extended region in the
$(L_\nu\langle E_\nu^2\rangle)_\mathrm{corr}$-$(\dot{M}M_\mathrm{NS})^{3/5}$-plane, 
we additionally present 2D and 3D simulations of more massive, black-hole forming progenitors. These models from a
forthcoming paper (Summa et al., in prep.) possess higher values of the ``heating functional'' $L_\nu\langle E_\nu^2\rangle$, 
as well as larger values for the mass-accretion rate $\dot{M}$ and neutron star mass $M_\mathrm{NS}$.
With the large set of 2D and 3D models at hand, we show that the generalized critical condition 
provides an excellent criterion to separate models with failed shock expansion from those with successful
shock revival.

The paper is structured as follows. After a brief description of the numerical setup in Sect.\,\ref{sec:num},
our simulation results are presented in Sect.\,\ref{sec:res}. In Sect.\,\ref{sec:crit}, we place our model set 
within the context of the generalized critical luminosity condition introduced by \citet{Summa2016} and conclude
in Sect.\,\ref{sec:con}.

\begin{deluxetable*}{lccrcc}
\tablecaption{Overview of 2D and 3D simulations \label{tab:table}}
\tablehead{
\colhead{model} & \colhead{dimension} & \colhead{rotation} & \colhead{angular resolution} & \colhead{explosion time\tablenotemark{a}} & \colhead{end of simulation}
}
\startdata
m15\_3D\_artrot\_2deg & 3D & fast & 2 degrees & 0.217\,s & 0.460\,s \\
m15\_3D\_rot\_2deg & 3D & \citet{Heger2005} & 2 degrees & - & 0.368\,s \\
m15\_3D\_rot\_6deg & 3D & \citet{Heger2005} & 6 degrees & - & 0.453\,s \\
m15\_3D\_norot\_4deg & 3D & no & 4 degrees & - & 0.329\,s \\
m15\_3D\_norot\_6deg & 3D & no & 6 degrees & - & 0.377\,s \\
\hline
m15\_2D\_artrot\_1.4deg & 2D & fast & 1.4 degrees & - & 0.434\,s \\
m15\_2D\_artrot\_2deg & 2D & fast & 2 degrees & - & 0.476\,s \\
m15\_2D\_rot\_1.4deg & 2D & \citet{Heger2005} & 1.4 degrees & 0.398\,s & 0.521\,s \\
m15\_2D\_rot\_2deg & 2D & \citet{Heger2005} & 2 degrees & 0.416\,s & 0.500\,s \\
m15\_2D\_norot\_1.4deg & 2D & no & 1.4 degrees & 0.454\,s & 0.567\,s \\
m15\_2D\_norot\_2deg & 2D & no & 2 degrees & 0.513\,s & 0.626\,s \\
\enddata
\tablenotetext{a}{Onset of explosion defined by the point in time when the ratio of advection to heating time scale reaches unity.}
\end{deluxetable*}

\section{Numerical Setup} \label{sec:num}

The calculations presented in this paper were performed with the neutrino-hydrodynamics
code \textsc{Prometheus-Vertex}. This tool couples the hydrodynamics solver \textsc{Prometheus} 
\citep{Fryxell1989} via lepton number, energy, and momentum source terms with 
the neutrino transport module \textsc{Vertex} \citep{Rampp2002b}. Applying three-flavor,
energy-dependent, ray-by-ray-plus (RbR+) neutrino transport \citep[see][]{Buras2006a}, \textsc{Vertex} 
includes the full set of neutrino reactions and microphysics currently available 
\citep[cf.][]{Marek2009a,Mueller2012c}. At high energies, the equation of state (EoS)
of \citet{Lattimer1991} with a nuclear incompressibility of 220\,MeV was used. 
We applied a low-density EoS with 23 nuclear species for nuclear statistical equilibrium (NSE) in regions
with temperatures/densities above/below a certain threshold value which was chosen differently
before and after bounce. Below NSE temperature, we used the flashing treatment as an approximation
for nuclear burning \citep[cf.][]{Rampp2002b}.
The simulations were conducted with a 1D gravitational potential including general
relativistic corrections as described for Case A in \citet{Marek2006a}. Since rotational deformation
of the proto-neutron star can only be observed at late times
for the fastest rotating progenitor model of this study (see discussion below), the application of 
a 1D potential as well as the RbR+ method is reasonably well justified to follow our models during
the post-bounce accretion phase towards the onset of explosion.

\begin{figure*}
\centering
\includegraphics[width=\textwidth]{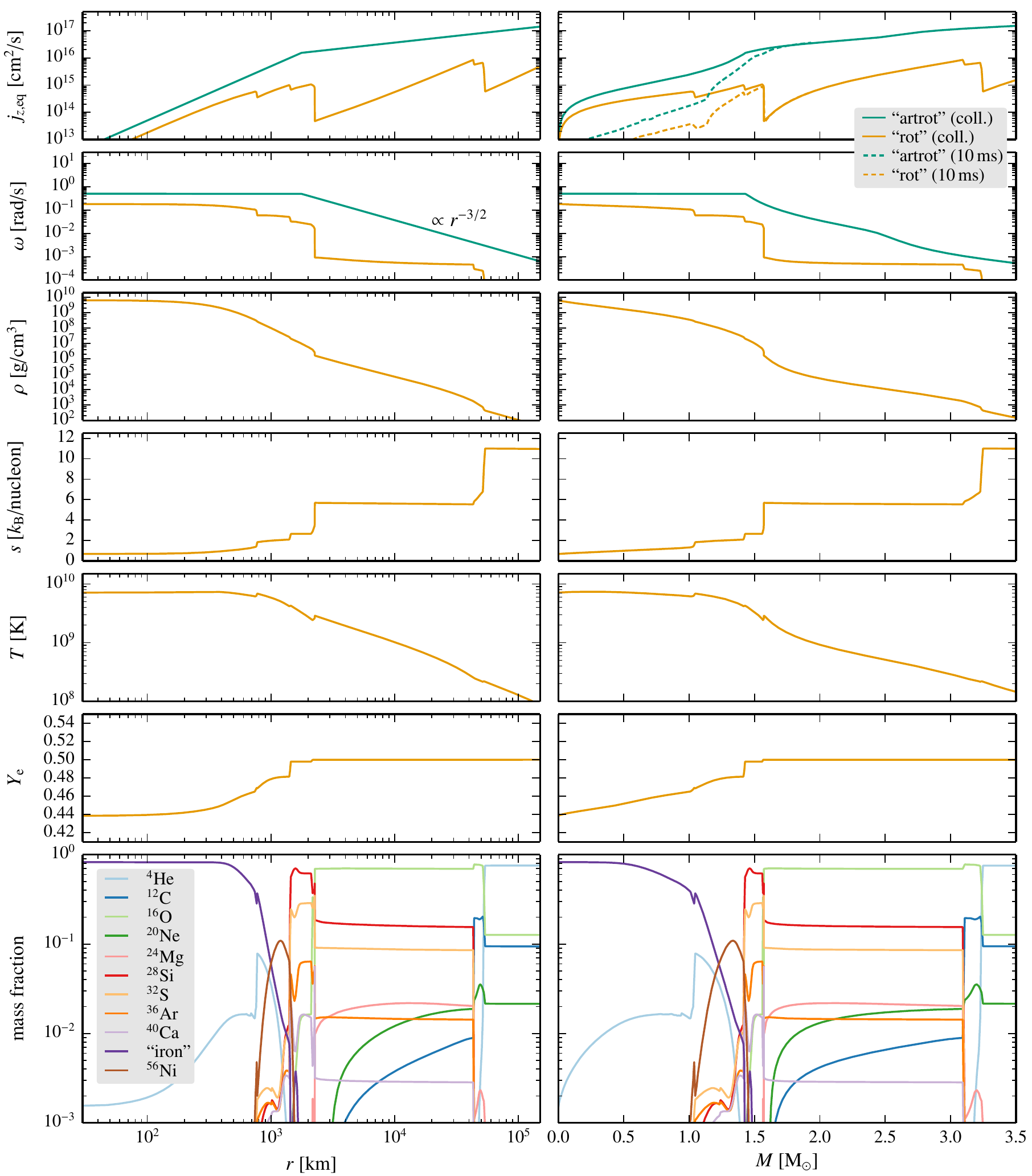}
\caption{Initial profiles of angular momentum, angular velocity, density, entropy, temperature, electron fraction, 
and nuclear mass fractions (from top to bottom) versus radius (left column) and mass (right column) as given by the 
1D stellar evolution calculations of \citet{Heger2005} for their model m15u6 at the onset of collapse (case ``rot''). In addition,
the enhanced angular momenta and velocities for our ``artrot'' case are shown (green lines in the top four panels). The difference between initial
(i.e., pre-collapse; lines denoted by ``coll.'' in the legend box)
and 10\,ms-post-bounce values (dashed lines in the upper right panel; cases denoted by ``10\,ms'' in the legend box) visualizes our erroneous loss of 
angular momentum during the collapse phase (see Sect.\,\ref{sec:num}). For the mass fractions, the notation ``iron''
refers to the sum of neutron-rich isotopes in the iron group, in particular $^{54}$Fe, $^{56}$Fe, and $^{58}$Fe.\label{omega}}
\end{figure*}

\begin{figure*}
\centering
\includegraphics[width=0.85\textwidth]{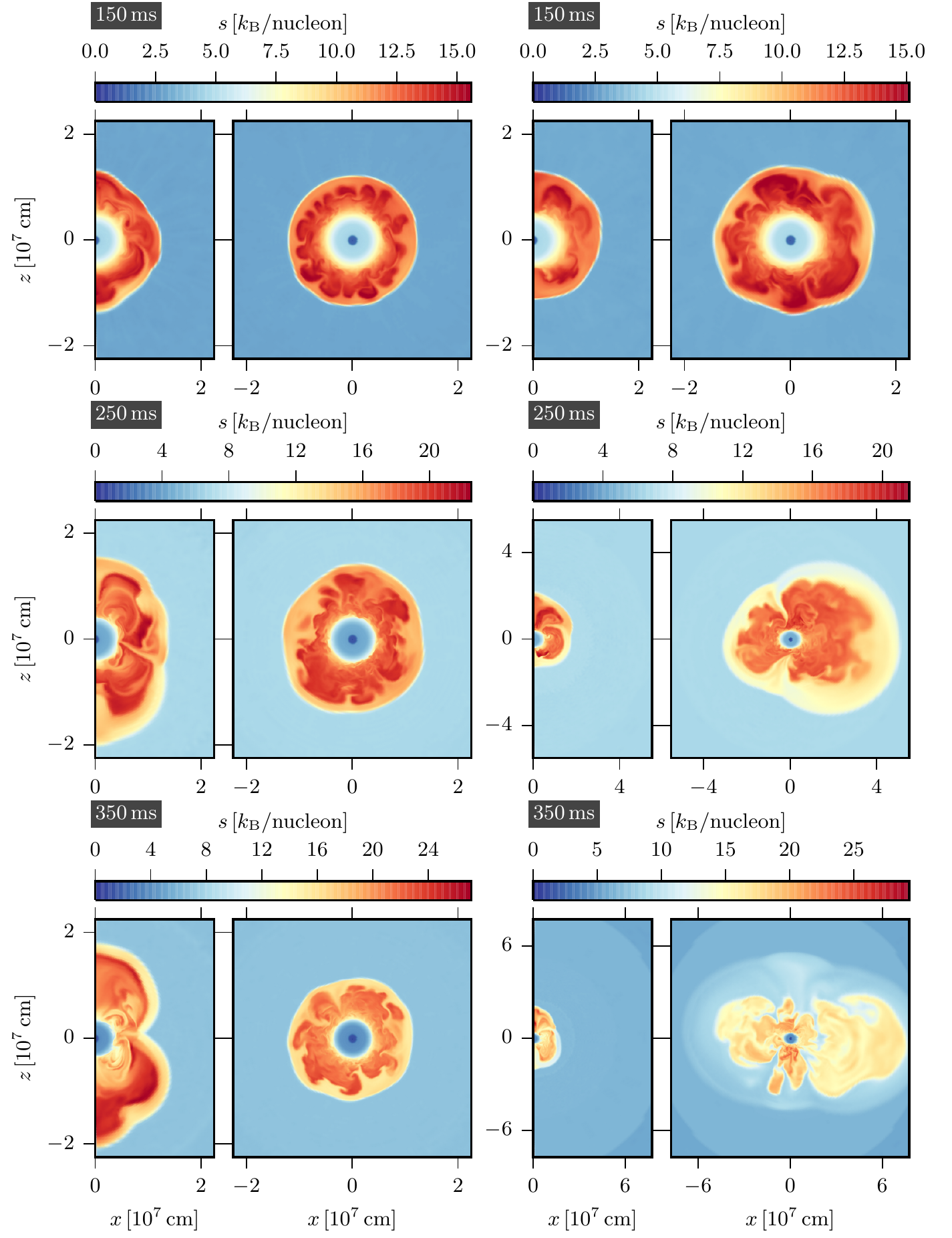}
\caption{Snapshots of entropy for simulations with moderate (left column, exploding model m15\_2D\_rot\_2deg and nonexploding model m15\_3D\_rot\_2deg) 
and fast rotation (right column, nonexploding model m15\_2D\_artrot\_2deg and exploding model m15\_3D\_artrot\_2deg). 
From top to bottom, cross-sectional cuts in the $x$-$z$-plane (the $z$-axis coincides with the rotation axis of the progenitor model)
 at 150\,ms, 250\,ms, and 350\,ms after core bounce are shown. In the left panels, the 2D models are depicted, 
the 3D simulations are displayed in the right panels. Note the changing scales of the plots in the
right column.\label{entropy_cuts}}
\end{figure*}

For all simulations, we used the (spherically symmetric) 15\,M$_\odot$ progenitor model 
m15u6\footnote{\url{http://www.2sn.org/stellarevolution/magnet/}} from the stellar evolution
calculations of \citet{Heger2005}, which include angular momentum transport by magnetic fields.
Up to 10\,ms after bounce, all models were computed on a spherical polar grid in 2D with initially
400 radial and 96 or 128 angular zones (denoted with ``2deg'' or ``1.4deg'' in the model name).
Computing the infall phase until 10\,ms post bounce in 3D is not necessary. The reason is
that triaxial instabilities are not expected to occur until this time for the moderately fast
pre-collapse rotation associated with the angular momentum profiles adopted for this work.
Indeed, our simulations confirmed that the
first hydrodynamical flows with relevant 3D effects do not develop before 100\,ms after bounce.
We considered two cases for the initial rotation profiles that are indicated in the model names: 
While the abbreviation ``rot'' denotes simulations using the initial angular velocities given by \citet{Heger2005}
(corresponding to an equatorial rotation velocity of 300\,km\,s$^{-1}$ on the zero-age main sequence),
simulations with the designation ``artrot'' employed a rotation profile that was already used in 
previous publications by \citet{Mueller2004,Buras2006a,Marek2009a}. Besides allowing us to 
refer to our experience with the previous calculations, this profile, as before, was chosen to
approximately reproduce the overall trends of pre-collapse models found in stellar evolution
models with rotation, namely a roughly rigidly rotating iron core surrounded by shells with 
radially decreasing angular velocity but increasing angular momentum for a rotationally stable 
stratification. Moreover, the adopted ``artrot'' profile maximizes rotational effects
but at the same time avoids sub-millisecond periods of the forming proto-neutron star.
Furthermore, our choice guarantees that the rotational deformation of the stellar 
core is negligible at early times after bounce \citep[see][]{Mueller2004}. 
For comparison, both initial 
angular velocity profiles at the beginning of collapse are shown in Fig.\,\ref{omega}.
In both cases, the angular velocity was assumed to be constant on spheres.

Because of an accidental mistreatment of fictious forces in the 2D simulations up to 10\,ms after 
bounce\footnote{Several code lines had been deactivated for test calculations and reactivation had been forgotten
afterwards.}, the angular momentum of the infalling material is initially not exactly conserved. This leads to 
slower rotation frequencies of shells that have collapsed significantly up to core bounce and a corresponding reduction
of the angular momentum mostly of matter that gets integrated into the proto-neutron star during the later accretion 
(see also Fig.\,\ref{omega}). However, because this deficit only holds for the initial collapse phase in 2D
and hardly affects the material accreted at times later than 100\,ms of post-bounce evolution, it is not crucial for 
our discussion of the angular momentum evolution of the gain region and the resulting effects on the explosion physics (see Sect.\,\ref{sec:res}).

The growth of aspherical instabilities was seeded by randomly imposed cell-to-cell perturbations 
of 0.1\,\% in radial velocity on the entire computational grid at the beginning of the simulation.
In the case of the 3D simulations, at 10\,ms after bounce the data were mapped to 
an axis-free Yin-Yang grid \citep{Kageyama2004,Wongwathanarat2010} with
an angular resolution as given in the model name. In order to avoid excessive time step
limitations at the grid center, the innermost 1.6\,km of the stellar core were treated in spherical
symmetry. The zones of the radial grid were non-equidistantly distributed from the center with
a reflecting boundary at the coordinate origin to an outer boundary of $10^9$\,cm with an inflow
condition. During the simulations, the radial grid was gradually refined, reaching a number of 
$\sim600$ zones and a resolution of $\Delta r/r \sim 3.5\times 10^{-3}$ at the gain radius at the
time the simulations were stopped. For the neutrino transport, 12 geometrically spaced energy bins 
with an upper bound of 380\,MeV were employed. 

For one 3D simulation up to $\sim\,$0.5\,s after bounce with an angular resolution of two degrees  
(e.g.\ model m15\_3D\_artrot\_2deg), the computational costs amounted to roughly 50 million CPU hours 
on, for example, {\em Sandy Bridge-EP Xeon E5-2680 8C} processors
on SuperMUC of the Leibniz Supercomputing Centre
(LRZ). Typically, roughly 16,000 cores were employed in parallel. For the whole model set,
the computational costs summed up to about 130 million CPU hours.

\section{Results and Discussion} \label{sec:res}

\begin{figure}
\includegraphics[width=\columnwidth]{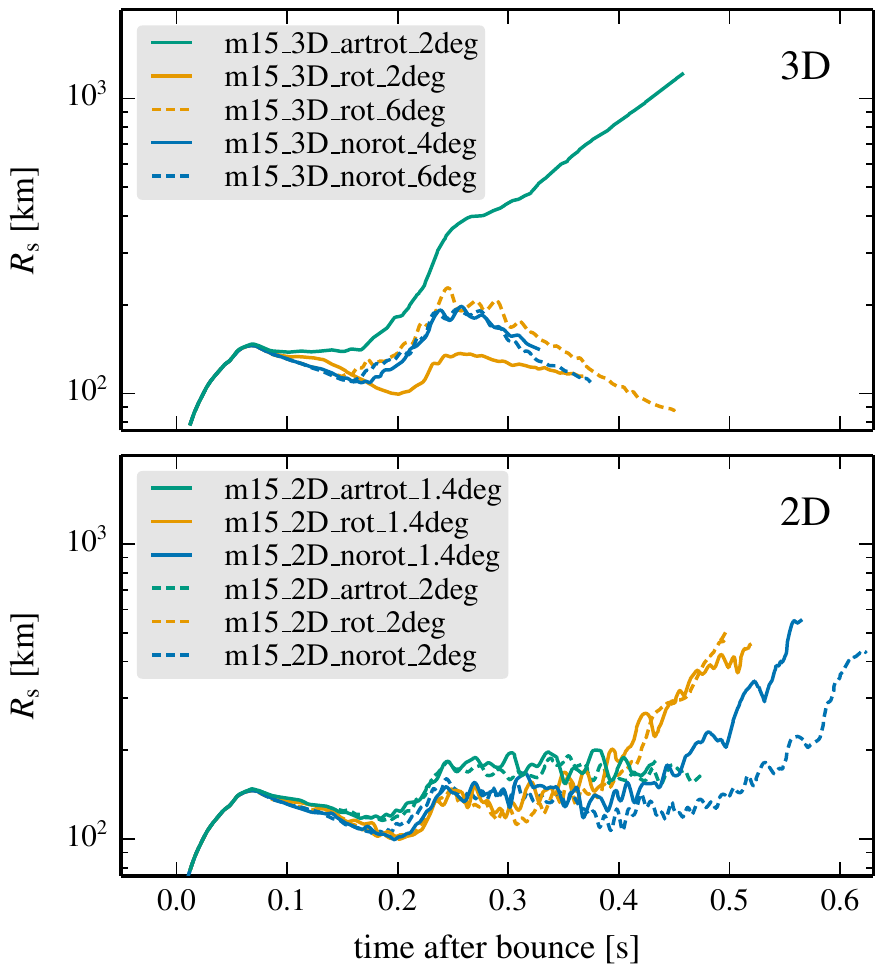}
\caption{Time evolution of shock radius (averaged over all angular directions)  
for our simulations in 3D (upper panel) and in 2D (lower panel). 
The curves are smoothed by running averages of 5\,ms.\label{shock_radii}}
\end{figure}

\begin{figure*}
\includegraphics[width=\textwidth]{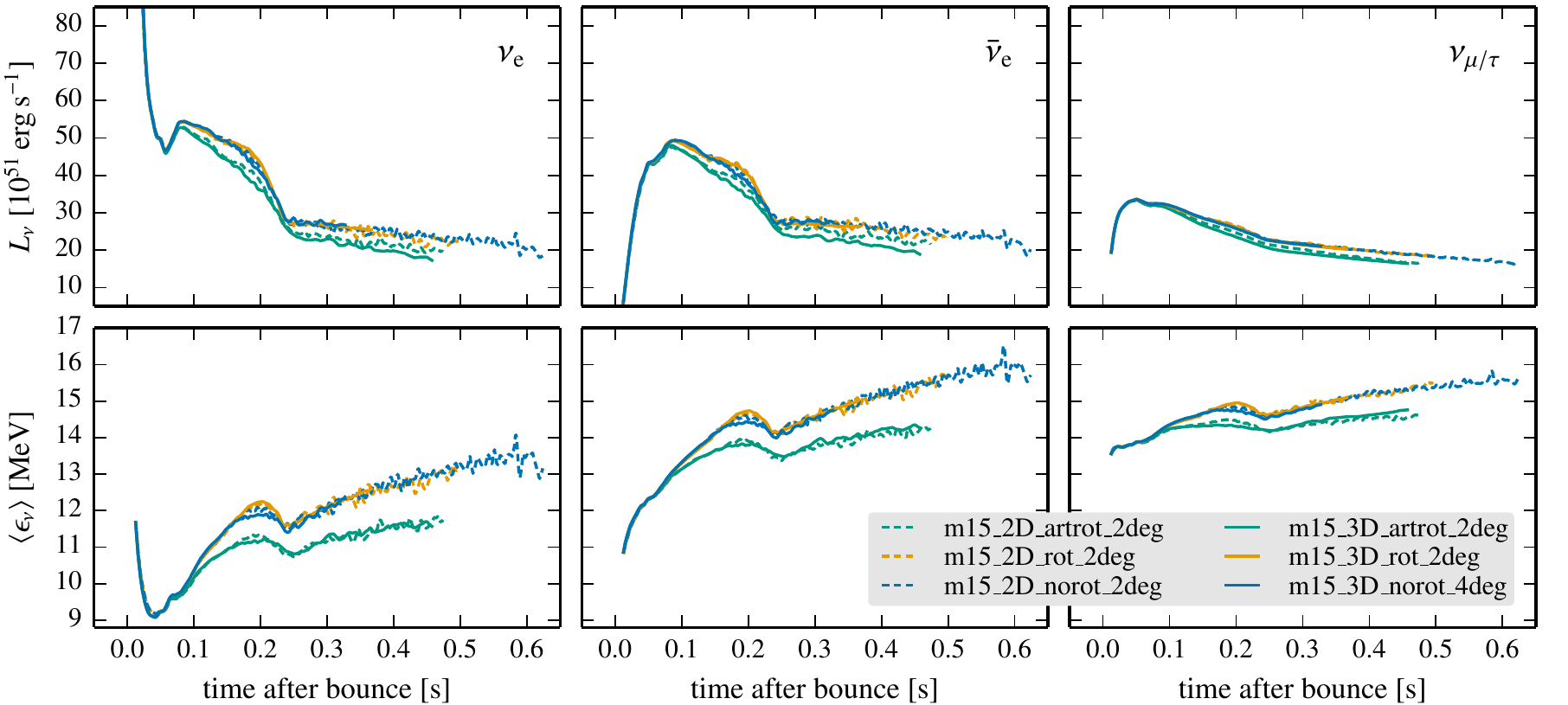}
\caption{Time evolution of neutrino luminosities (first row) and neutrino mean 
energies (ratio of neutrino energy flux to neutrino number flux; second row) for the simulations in 2D (dashed lines) and in 3D (solid lines; 
lab-frame quantities, evaluated at 400\,km and given for an observer
at infinity). From left to right, the angular averages are shown 
for $\nu_\mathrm{e}$, $\bar{\nu}_\mathrm{e}$, and $\nu_{\mu/\tau}$. The curves are smoothed by running averages of 
5\,ms.\label{nu_prop}}
\end{figure*}

In this section, we present the results of our 2D and 3D CCSN simulations of
rotating progenitor stars. After a short overview of the complete model set
and a comparison between rotational effects in 2D and in 3D, we focus on our
two best resolved 3D models with moderate and enhanced rotation rates and 
discuss the influence of rotation on the explosion physics in detail.

\subsection{General Properties of Rotation-supported Models in 2D and 3D}

\begin{figure}
\includegraphics[width=\columnwidth]{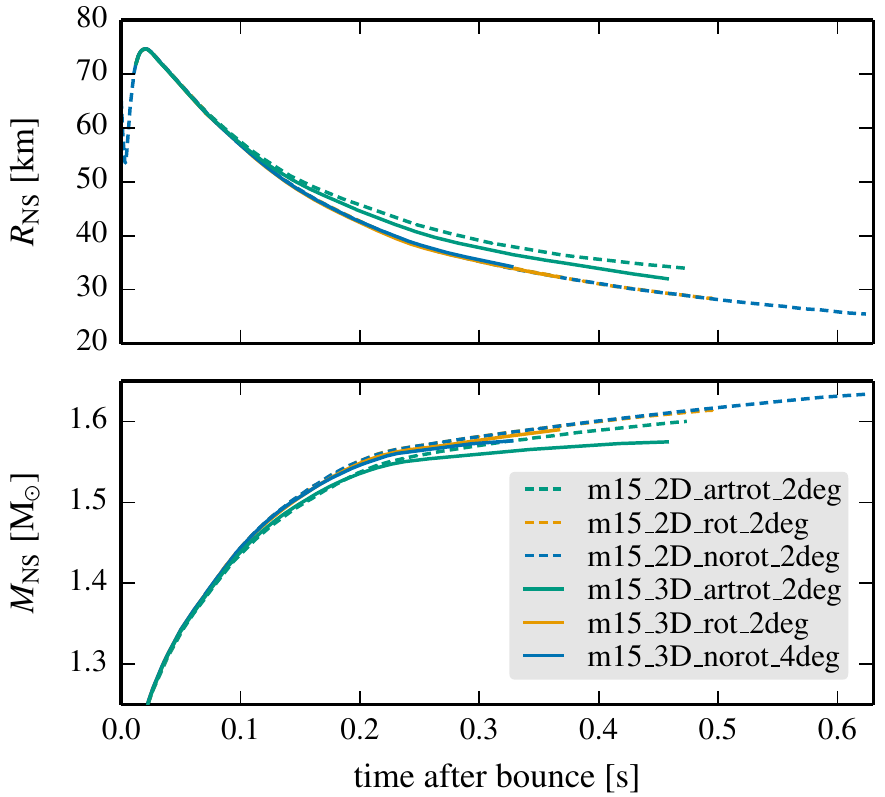}
\caption{Time evolution of the average neutron star radius (upper panel) and baryonic neutron star mass (lower panel).
Both quantities are defined by the radius where the density is $10^{11}\,\mathrm{g}\,\mathrm{cm}^{-3}$ 
on the angle-averaged density profile. The curves are smoothed by running averages of 5\,ms.\label{ns_radii}}
\end{figure}

\begin{figure}
\includegraphics[width=\columnwidth]{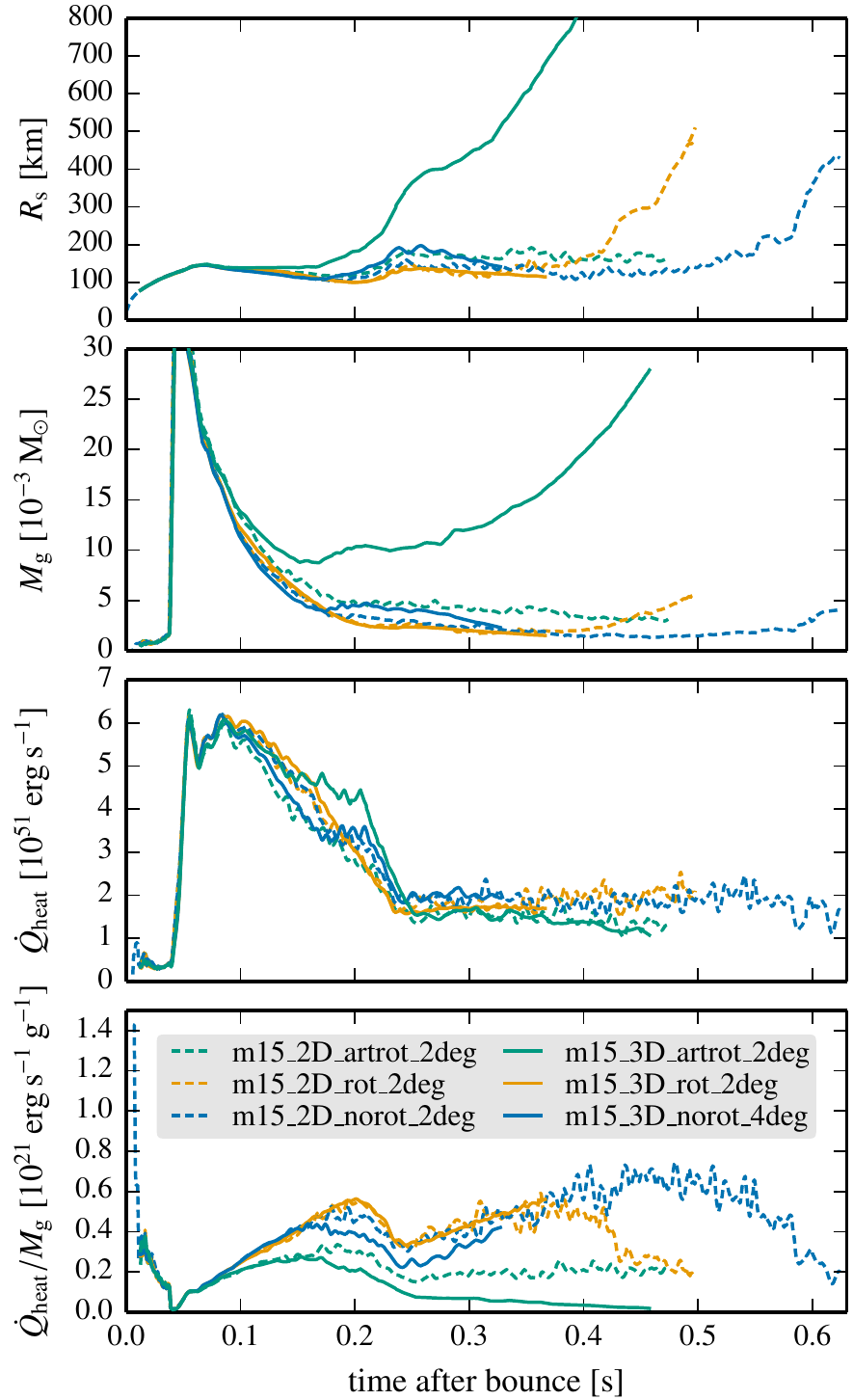}
\caption{Time evolution of average shock radius, gain layer mass, neutrino heating rate,
and neutrino heating rate per unit mass in the gain layer (from top to bottom). 
The curves are smoothed by running averages of 
5\,ms.\label{gain_quants}}
\end{figure}

\begin{figure}
\includegraphics[width=\columnwidth]{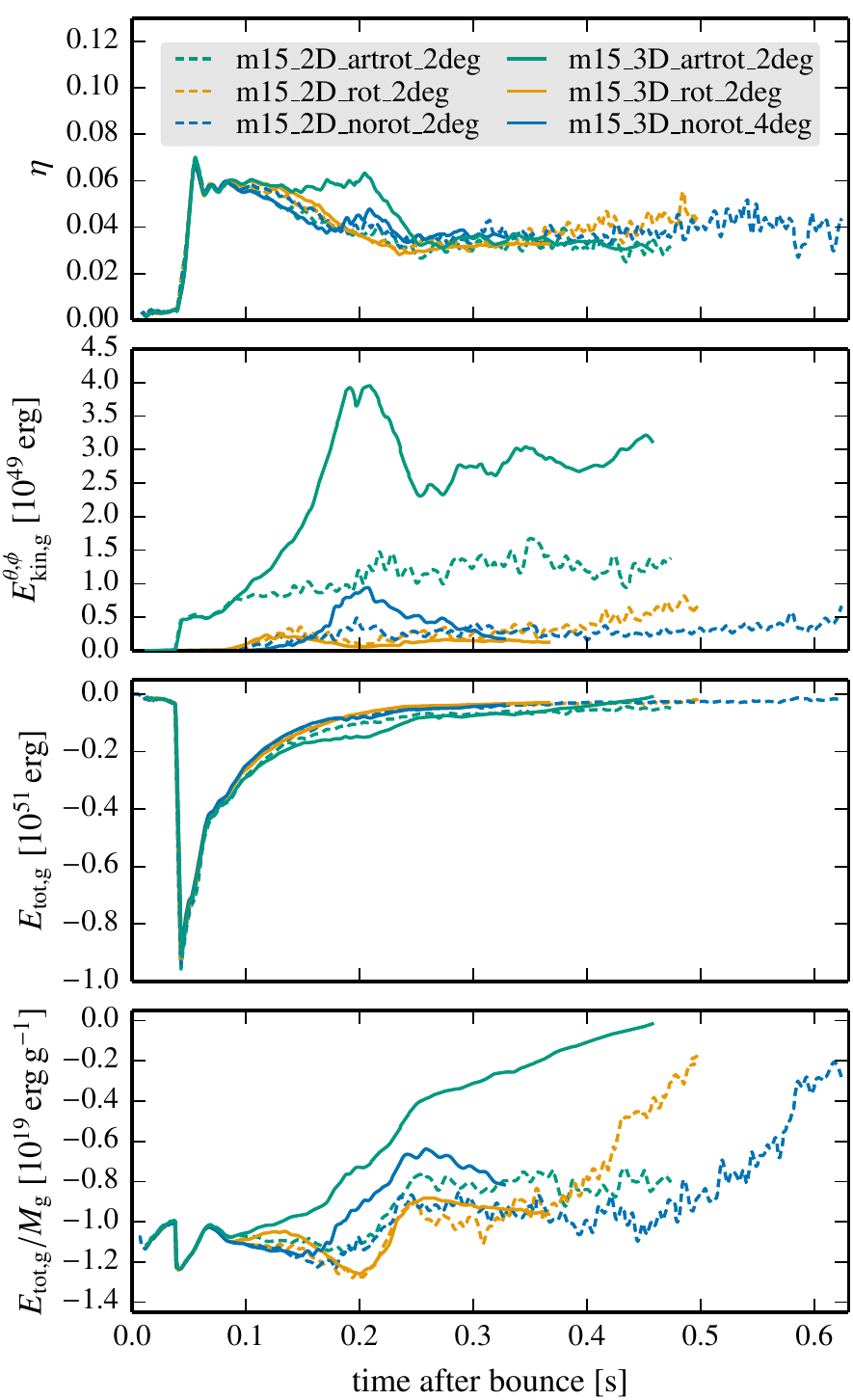}
\caption{Time evolution of neutrino heating efficiency, 
kinetic energy of nonradial motions, total energy, and total specific energy in the gain layer (from top to bottom). 
The curves are smoothed by running averages of 
5\,ms.\label{gain_ene}}
\end{figure}

\begin{figure}
\includegraphics[width=\columnwidth]{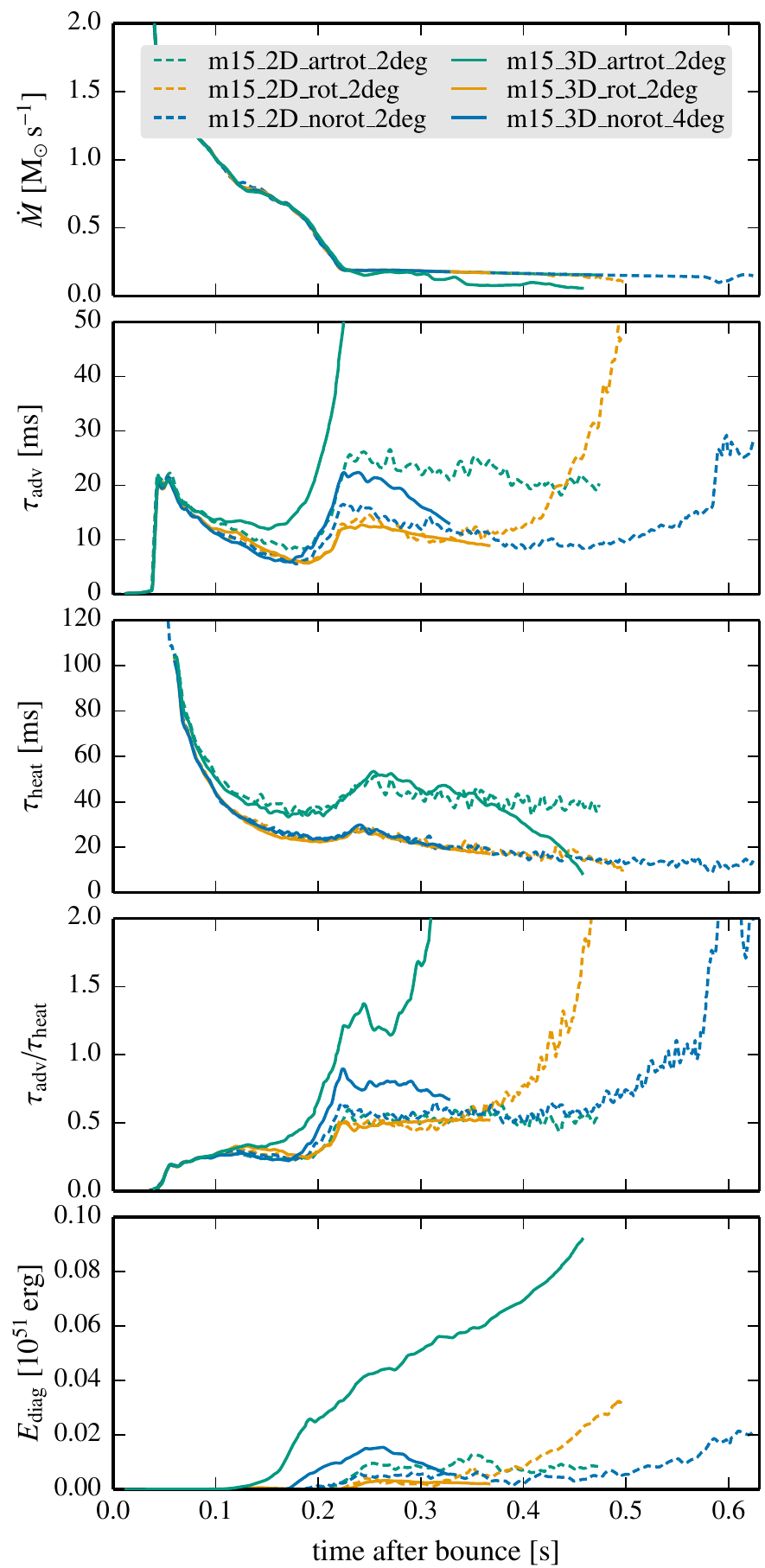}
\caption{Time evolution of mass-accretion rate (evaluated at a radius of 400\,km), 
advection and heating time scale in the gain layer as well as the ratio of the two
time scales, and diagnostic explosion energy (from top to bottom). The curves are smoothed by running averages of 5\,ms.\label{time_scales}}
\end{figure}

\begin{figure*}
\centering
\includegraphics[width=0.95\textwidth]{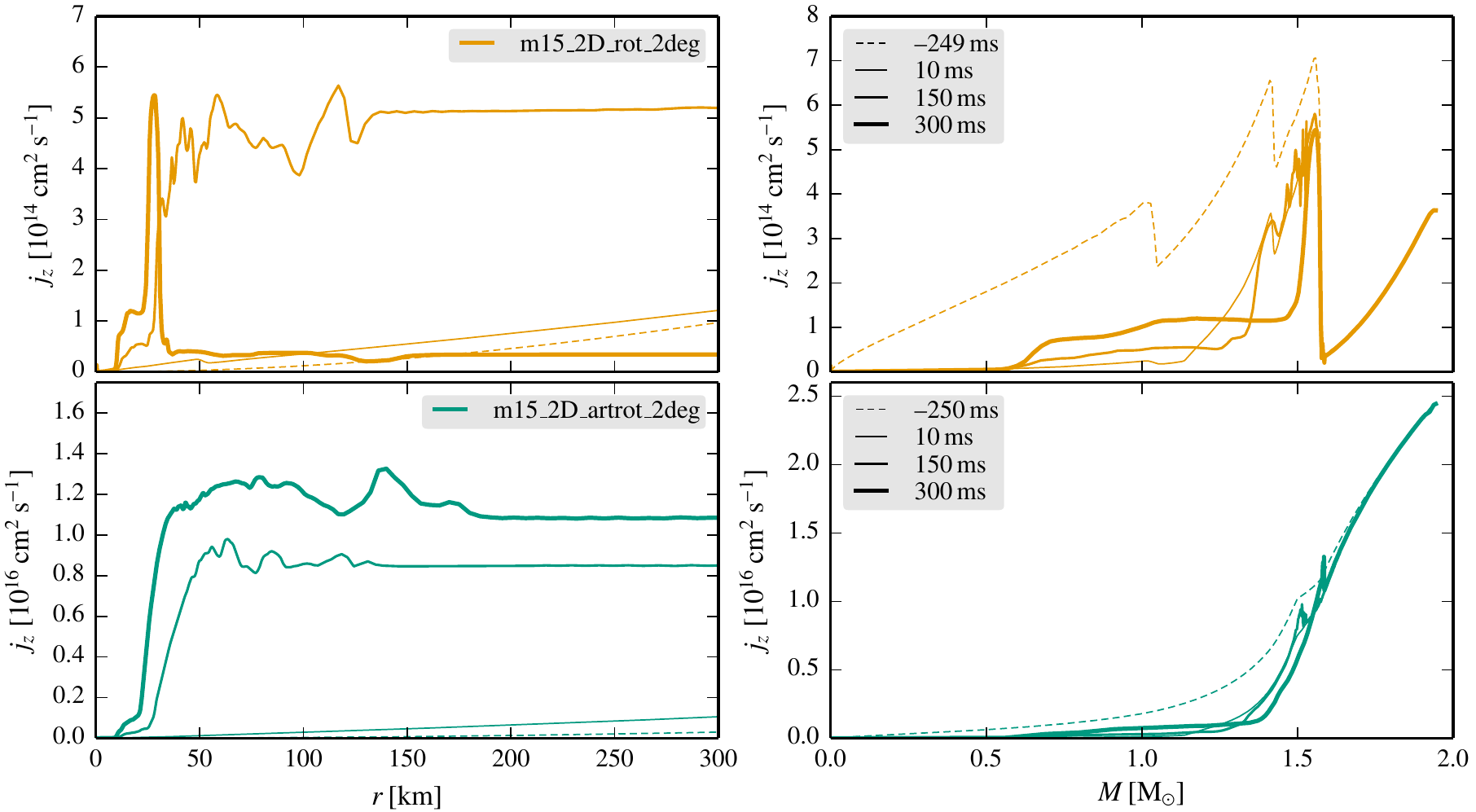}
\caption{Time evolution of angular momentum profiles vs. radius (left) and enclosed mass (right) for the moderately
(upper row) and fast rotating cases (lower row) in 2D. The enclosed mass $M$ is defined as the mass encompassed by spheres of 
growing radius (and is therefore not an ideal Langrangian coordinate). The specific angular momentum $j_z$ is computed as the 
ratio of the $z$-component of the angular momentum in spherical shells divided by the mass of the shells. 
Note that an accidental mistreatment of fictious forces in the
simulation code up to 10\,ms after bounce leads to a temporary violation of angular momentum conservation, but this only affects
the final neutron star spins.\label{ang_prop}}
\end{figure*}

\begin{figure*}
\centering
\includegraphics[width=0.95\textwidth]{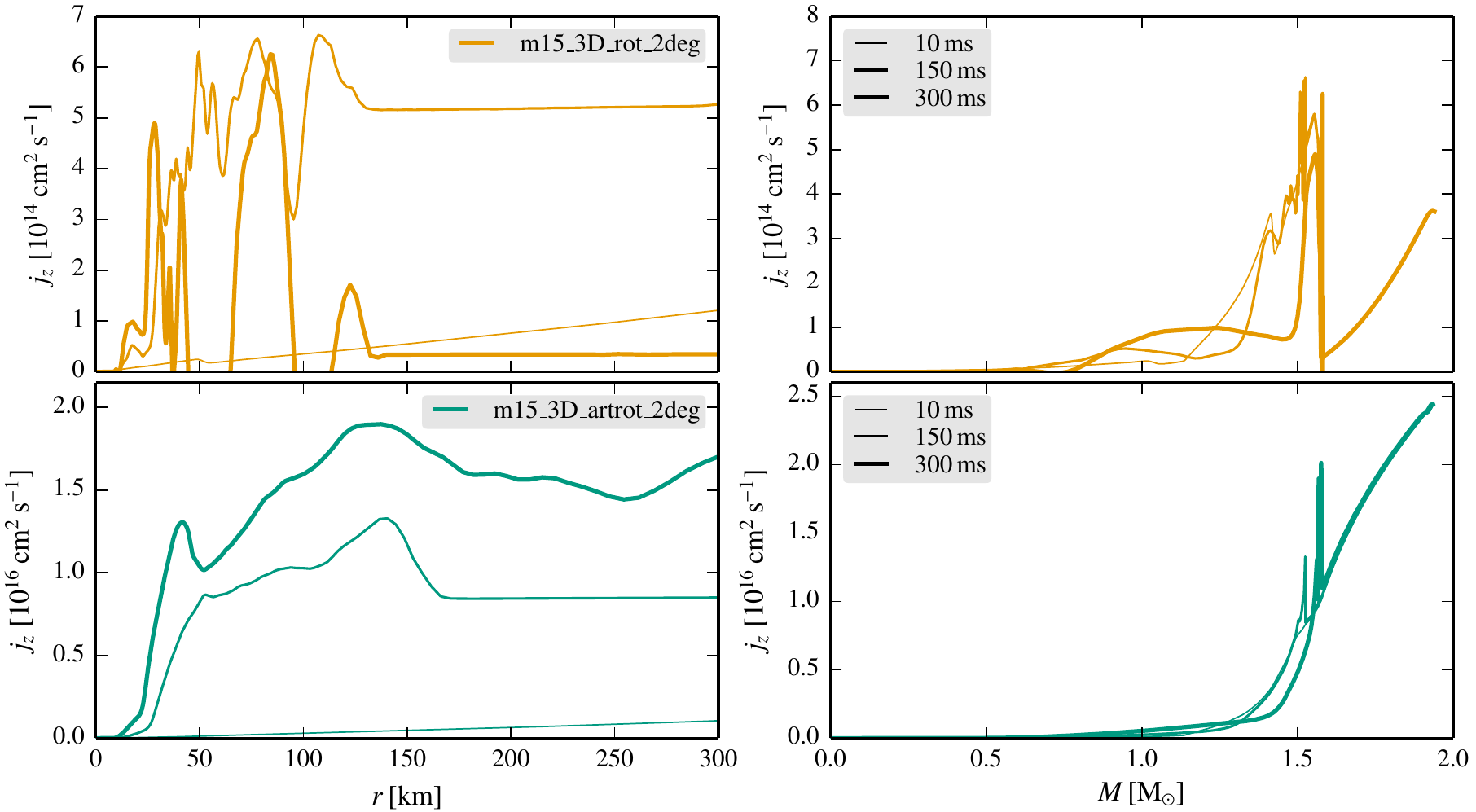}
\caption{Time evolution of angular momentum profiles vs. radius (left) and enclosed mass (right) for the moderately
(upper row) and fast rotating cases (lower row) in 3D. Definition of quantities analogous to Fig.\,\ref{ang_prop}.\label{ang_prop_3D}}
\end{figure*}

In Fig.\,\ref{shock_radii}, the time evolution of the mean shock radius is given
for the whole set of five 3D and six 2D models. Table\,\ref{tab:table} provides an overview of the computed
2D and 3D models. The 3D model set consists of
a fast rotating model (angular resolution of 2 degrees, for more details see
Sect.\,\ref{sec:num}), two moderately rotating models (angular resolutions of 2 and 6 degrees),
and two nonrotating models (4 and 6 degrees). Only the fast rotating
model m15\_3D\_artrot\_2deg develops an explosion, all other 3D models do not explode before  
the simulations are stopped. The larger shock radii of the less resolved models between
0.2 and 0.3\,s after bounce compared to model m15\_3D\_rot\_2deg are due to a more pronounced activity of the standing 
accretion-shock instability (SASI), which is not strongly present in the better resolved model with a moderate rotation rate (see also Sect.\,\ref{sub:3D}).
In the models with angular cell sizes of 4 or 6 degrees, turbulent eddies due to convection cannot be resolved on small scales
and the growth of SASI modes seems to be facilitated by the suppression of parasitic Rayleigh-Taylor and Kelvin-Helmholtz 
instabilities. This is why we will focus on the 2-degree-runs in the discussion of the results
and will only depict the results of the three best-resolved 3D simulations in all subsequent figures.

The 2D model set consists of six models with angular resolutions of 1.4 and 2 degrees for each of
the three different (fast, moderately, and nonrotating) cases. Since the time evolution of the 
better resolved models is very similar to the models with an angular resolution of 2 degrees, 
only the latter ones are presented in the following figures. This places the 2D and 3D simulations on
(roughly) equal footing with respect to resolution.

Compared to their 3D counterparts,
the behavior of the 2D models is different (see Figs.\,\ref{entropy_cuts} and \ref{shock_radii}). 
While the moderately and nonrotating 2D models explode,
the fast rotating 2D model does not evolve towards a successful explosion before the end of the simulation.
This distinct behavior underlines the need for simulations in three dimensions to assess the effects	
of rotation in the context of the neutrino-driven explosion mechanism.

A comparison of the neutrino luminosities and mean energies (see Fig.\,\ref{nu_prop}) shows only minor 
differences between the 2D (dashed lines) and 3D models (solid lines). While the effect of moderate rotation rates on the neutrino quantities 
is negligible (compare models m15\_2D\_rot\_2deg and m15\_3D\_rot\_2deg to the nonrotating cases), 
the influence of rotation is clearly visible in models m15\_2D\_artrot\_2deg and m15\_3D\_artrot\_2deg. After 100\,ms
post bounce, neutrino luminosities and mean energies are significantly reduced due to the more extended and 
cooler rotating neutron star \citep[see Fig.\,\ref{ns_radii} and][]{Buras2006a,Marek2009a}. In the 3D model the effect
is even stronger than in the corresponding 2D case due to the growth of the shock stagnation radius, which reduces the 
mass-accretion rate of the neutron star (see Fig.\,\ref{ns_radii} for a slower growth of the neutron star mass)
and its $\nu_\mathrm{e}$ and $\bar\nu_\mathrm{e}$ emission, in particular, since energy is stored in rotation associated with spiral
SASI mass motions instead of being released by neutrinos.

Therefore, the neutrino heating rate per unit mass, $\dot{Q}_\mathrm{heat}/M_\mathrm{g}$, is considerably lower than in the
moderately and nonrotating models (see Fig.\,\ref{gain_quants}). 
Here, the neutrino heating rate is defined as the integral of the net neutrino energy
deposition rate per volume $q_\mathrm{e}$ (heating minus cooling) over the gain layer,
\begin{equation}
\dot{Q}_\mathrm{heat}=\int 
\limits_{V_\mathrm{g}}q_\mathrm{e}\,\mathrm{d}V,
\end{equation}
and $M_\mathrm{g}$ is given by the mass enclosed in the gain layer,
\begin{equation}
M_\mathrm{g}=\int \limits_{V_\mathrm{g}}\rho\,\mathrm{d}V.
\end{equation}
Because shock expansion is supported by the spiral SASI mode in model m15\_3D\_artrot\_2deg, 
the mass in the gain layer grows and overcompensates for the decrease of the specific heating rate. 
For this reason the heating rate $\dot{Q}_\mathrm{heat}$ and the heating efficiency, defined by the ratio of the total 
energy deposition rate to the sum of the radiated electron neutrino and electron 
antineutrino luminosities,
\begin{equation}
\eta=\frac{\dot{Q}_\mathrm{heat}}{L_{\nu_\mathrm{e}}+L_{\bar{\nu}_\mathrm{e}}},
\end{equation}
are even highest in model m15\_3D\_artrot\_2deg (see Figs.\,\ref{gain_quants} and \ref{gain_ene}).

This also leads to a longer neutrino heating time scale (see Fig.\,\ref{time_scales}),
\begin{equation}
\tau_\mathrm{heat}=\frac{\left|E_\mathrm{tot,g}\right|}{\dot{Q}_\mathrm{heat}},
\label{eq:theat}
\end{equation}
where $E_\mathrm{tot,g}$ is the total (i.e., internal plus kinetic (including rotational) plus gravitational) 
energy in the gain layer. Due to the larger mass contained in the gain layer, the value 
of the total energy is lower (more negative), but the specific total energy in the gain layer is significantly higher than in the other models
because of the additional kinetic energy provided by the spiral SASI mode (see Fig.\,\ref{gain_ene}).

In 2D, however, all models show roughly the same time-scale ratio $\tau_\mathrm{adv}/\tau_\mathrm{heat}$ until
models m15\_2D\_norot\_2deg and m15\_2D\_rot\_2deg finally explode.
In the fast rotating model m15\_2D\_artrot\_2deg, centrifugal forces stabilize the accretion shock at larger
radii (see Fig.\,\ref{shock_radii} and Fig.\,\ref{gain_quants}, upper panel) and increase the advection time scale noticeably (compare the 
advection time scales of all 2D models (dashed lines) between 200 and 400\,ms after bounce in Fig.\,\ref{time_scales}, 
second panel from top). Here, the advection time scale is defined by the dwell time of matter in the gain region 
\citep[cf.][]{Buras2006b,Marek2009a}:
\begin{equation}
\tau_\mathrm{adv}\equiv\tau_\mathrm{dwell}\approx\frac{M_\mathrm{g}}{\dot{M}},
\label{eq:dwell}
\end{equation}
where $\dot{M}$ is the mass-accretion rate through the shock. But this favorable influence of rotation on the explosion
conditions cannot overrule the reduced energy deposition by neutrinos due to the lower neutrino luminosities and mean energies in model m15\_2D\_artrot\_2deg.
Different from model m15\_2D\_rot\_2deg and model m15\_2D\_norot\_2deg, this fast rotating model does not explode until the simulation is stopped.

In contrast, in 3D \textit{only} the setup with the fastest rotation (model m15\_3D\_artrot\_2deg) develops an explosion.
The nonrotating 3D case (model m15\_3D\_norot\_4deg) does not explode although it exhibits a more favorable time-scale ratio
(see Fig.\,\ref{time_scales}) than the exploding nonrotating and rotating 2D models m15\_2D\_norot\_2deg and m15\_2D\_rot\_2deg.
This finding is consistent with previous results for nonrotating 11.2, 20, and 27\,M$_\odot$ progenitors, which also
did not explode in 3D whereas their 2D counterparts did \citep[see][]{Hanke2013a,Tamborra2014,Tamborra2014a}.

The fast rotating model m15\_3D\_artrot\_2deg 
shows the largest values of the advection time scale shortly after bounce. This is caused by the early development
of a strong spiral SASI mode, which immediately drives the shock to larger radii and leads to a strong increase of 
the nonradial kinetic energy and the mass contained in the gain layer. This SASI spiral mode therefore has a much larger influence 
than SASI sloshing motions in model m15\_2D\_artrot 2deg (compare both models in 
Fig.\,\ref{gain_quants}). The explosion properties of these fast rotating models will be further discussed in Sect.\,\ref{sub:3D}.

\subsection{Angular Momentum Evolution}

The strong spiral SASI mode in model m15\_3D\_artrot\_2deg also leaves its imprint on the angular momentum
profiles: Compared to the 2D case, the angular velocities in the gain region are significantly increased 
(see lower left panels of Figs.\,\ref{ang_prop} and \ref{ang_prop_3D}).
Prominent spikes due to spiral-SASI activity can be seen in model m15\_3D\_artrot\_2deg at $M\approx1.5\,\mathrm{M}_\odot$ 
for $t=150\,\mathrm{ms}$ and at $M\approx1.6\,\mathrm{M}_\odot$ for $t=300\,\mathrm{ms}$ (see lower left panel of Fig.\,\ref{ang_prop_3D}).
Such a pronounced 2D-3D difference cannot be observed in the slow-rotating cases. Due to the strong decline of the angular velocities at
the Si/Si-O interface (see the sudden drop at a radius of $\sim$\,2000\,km in Fig.\,\ref{omega}), the angular momentum contained in the gain layer
decreases at later times (compare profiles for $r\gtrsim50\,\mathrm{km}$ at later and earlier times). While in 2D the specific angular
momentum of each mass element is preserved, in 3D the initially uniform, uni-directional rotation pattern is destroyed 
by convective and turbulent motions, i.e.\ the rotation rates of the accreted 
material are too low to dominate the fluid motions associated with hydrodynamic instabilities 
in the gain region in a global way (see the angular momentum profile at 300\,ms for 
model m15\_3D\_rot\_2deg in Fig.\,\ref{ang_prop_3D}).

\begin{figure}
\includegraphics[width=\columnwidth]{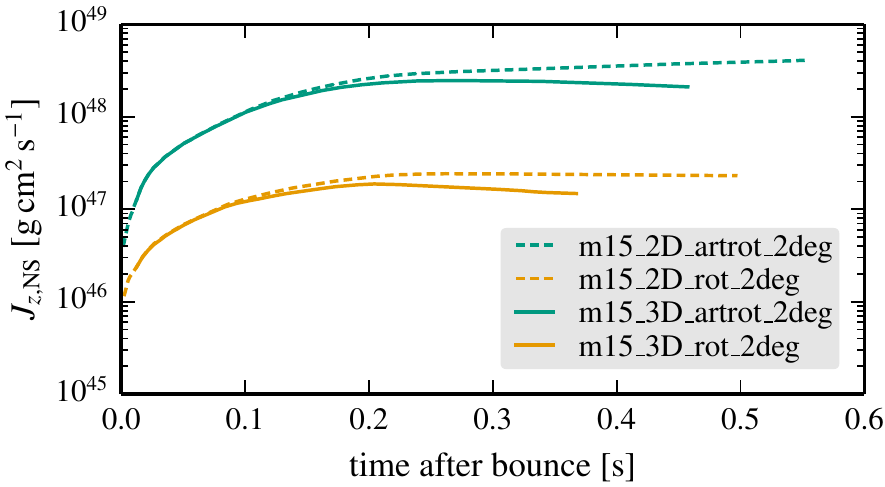}
\caption{Time evolution of the $z$-component of the angular momentum of the neutron star. The latter is defined by the 
mass at densities above $10^{11}\,\mathrm{g}\,\mathrm{cm}^{-3}$ for an angle-averaged density distribution.
The curves are smoothed by running averages of 5\,ms.\label{j_ns_z}}
\end{figure}

A comparison of the angular momentum contained in the neutron star in 2D and 3D (the $z$-axis is the rotation axis) is given in Fig.\,\ref{j_ns_z}
for the moderately and fast rotating cases. Angular momentum is naturally conserved in axisymmetry. Model m15\_2D\_rot\_2deg exhibits only a slow
growth when low-angular momentum material from outside the Si/Si-O interface is accreted, whereas $J_{z,\mathrm{NS}}$ for the more rapidly 
rotating model m15\_2D\_artrot\_2deg continues to increase as the neutron star mass grows with time by the infall of high-$j_z$ matter (Fig.\,\ref{ns_radii}).
In contrast, the 3D simulations show noticeable angular momentum losses towards later post-bounce times. Most of this effect is caused by the applied numerical schemes
which guarantee momentum and energy conservation, but do not conserve angular momentum up to machine precision. Especially at early
times in the linear regime of SASI growth, these losses are negligible and do not influence the interaction between rotation and the development
of SASI. 

A smaller part of the effect may be associated by the redistribution of low-angular momentum material into the neutron star (defined by $\rho\geq10^{11}\,\mathrm{g\,cm}^{-3}$)
due to nonspherical flows around the neutron star, while high-$j_z$ matter stays outside of the neutron star when the explosion sets in in 
model m15\_3D\_artrot\_2deg. Inside of the proto-neutron star, in 2D as well as in 3D, angular momentum is redistributed by convective
mass motions in the convective shell below the neutrinosphere. For this reason the radial profile of the specific angular momentum ($j_z$) 
versus mass in the region $M\lesssim1.5\,\mathrm{M}_\odot$ cannot be expected to be preserved (see right columns of Figs.\,\ref{ang_prop} and \ref{ang_prop_3D}).

The fact that the angular momentum of the neutron star is roughly one order of magnitude larger in the fast rotating models
directly translates into higher expected pulsar rotation rates. Assuming angular momentum conservation, a final neutron star radius of 12\,km, a moment of 
inertia of $0.35\,M_\mathrm{grav}\,R_\mathrm{NS}^2$ \citep[cf.][]{Lattimer2001} and taking the respective final values of the two 3D simulations into consideration, 
the final neutron star spin period evaluates to 5\,ms for model m15\_3D\_artrot\_2deg and to 66\,ms for model m15\_2D\_artrot\_2deg.
Note that the latter value is larger than the one derived in \citet{Heger2005} because of the erroneous angular momentum loss in our simulations 
during the collapse phase, which leads to a reduction of the angular momentum content in the inner core (see Sect.\,\ref{sec:num} and Figs.\,\ref{omega}, 
\ref{ang_prop}, and \ref{ang_prop_3D}).

\subsection{Turbulent Energy Cascades} \label{sub:turb}

\begin{figure}
\includegraphics[width=\columnwidth]{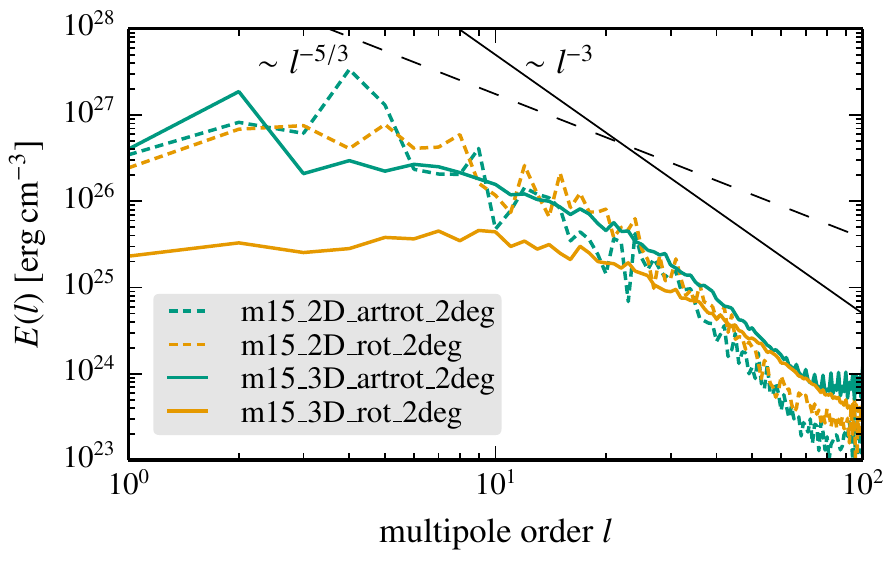}
\caption{Turbulent energy spectra $E(l)$ as functions of the multipole order $l$ 
for simulations in 3D (solid lines) and 2D (dashed lines). 
The spectra are based on a decomposition of the nonradial velocity components into
spherical harmonics at 150\,ms after bounce (before the onset of an explosion in model m15\_3D\_artrot\_2deg), 
averaged over a time interval of 5\,ms 
and a radius interval of 10\,km around the midpoint 
between gain radius and minimum shock radius. The thin dashed
and solid black lines indicate reference spectral slopes of $-5/3$ and $-3$.\label{turb}}
\end{figure}

In order to study the turbulent energy cascades in our models, we consider the energy spectrum $E(l)$ of turbulent
motions as a function of multipole order $l$ at a chosen radius in the gain region by using a decomposition of the nonradial
velocity field into spherical harmonics $Y_\mathrm{lm}(\theta,\phi)$ \citep[cf.][]{Hanke2012,Summa2016}:
\begin{equation}
E(l)= \sum\limits_{m=-l}^l\left|\int\limits_\Omega Y_{lm}^*(\theta,\phi)\sqrt{\rho}\,v_\mathrm{ang}\left(r,\theta,\phi\right)\mathrm{d}\Omega\right|^2,
\label{eq:sph_har}
\end{equation}
where $v_\mathrm{ang}=\sqrt{v_\theta(r,\theta,\phi)^2+v_\phi(r,\theta,\phi)^2}$. The results for the moderately
and fast rotating 2D and 3D models are shown in Fig.\,\ref{turb}. 

Due to the inverse energy cascade in axisymmetry,
the injected energy is not transferred to the dissipative range and $E(l)\propto l^{-5/3}$ approximately holds for
$l\lesssim30$ in models m15\_2D\_rot\_2deg and m15\_2D\_artrot\_2deg. At larger $l$, the direct vorticity
cascade leads to a power-law index of $\sim -3$ \citep{Kraichnan1967,Hanke2012}. Model m15\_3D\_rot\_2deg
develops a power-law spectrum following approximately $E(l)\propto l^{-5/3}$ at intermediate wavenumbers ($l\sim20-50$), which is 
characteristic for the energy transfer in 3D from large to small scales up to the dissipation range
at large $l$ \citep{Landau1959,Hanke2012}.

In contrast to the forward energy cascade of model m15\_3D\_rot\_2deg, model 
m15\_3D\_artrot\_2deg exhibits an energy spectrum which is very similar to the axisymmetric case. 
This unusual behavior can be explained by the presence of fast rotation. While the rotational effects in the 
moderately rotating 3D model are not strong enough to affect the energy transport, the global fast rotation of model 
m15\_3D\_artrot\_2deg leads to an anisotropic flow which approaches two-dimensionality
due to the rotational constraints of the flow geometry in the azimuthal direction. 

This phenomenon of rotational turbulence has also been studied in the context of geophysical and industrial flows and is subject to
current research in these fields \citep[see][and references therein]{Smith1999,Rubio2014,Godeferd2015}. Depending on the
Rossby number $\mathrm{Ro}$ being defined as the ratio of inertia force to Coriolis force \citep{Godeferd2015,Mueller2016},
\begin{equation}
\mathrm{Ro} = \frac{U}{2\omega L} \sim \frac{\left|v_r\right|}{2\left(R_\mathrm{s}-R_\mathrm{g}\right)\omega} \sim \frac{R_s^2}{2\tau_\mathrm{adv}j} \sim \frac{1}{4\pi} \frac{T_\mathrm{rot}}{\tau_\mathrm{adv}}
\end{equation}
($U$ and $L$ are characteristic velocity and length scales, $R_\mathrm{g}$ and $R_\mathrm{s}$ the mean gain and shock radius, 
and $\omega$, $j$, and $T_\mathrm{rot}$ the average angular velocity, specific angular momentum, and rotation period of the infalling shells), 
different regimes of turbulent dynamics can be identified. For $\mathrm{Ro}\gg1$, turbulent flows
evolve much faster than the rotation time scale and the turbulent structures are mostly unperturbed by rotation. This is the case for model m15\_3D\_rot\_2deg with a 
Rossby number of $\sim20$ in the gain layer at 150\,ms after bounce. Only at low $\mathrm{Ro}$, the rotation time scale is small enough for the Coriolis force
to effectively influence the flow dynamics. This can be observed in model m15\_3D\_artrot\_2deg where $\mathrm{Ro}$ evaluates to $\sim$0.5 at 150\,ms post bounce. 
The fact that the slopes of the energy spectra of the two 3D models become very similar for $l>50$ (although one has to note that the regime of large $l$ can suffer from resolution 
deficits) may be an indication that only large scales are affected by rotation in model m15\_3D\_artrot\_2deg, while small scales below 
the Zeman scale \citep[cf.][]{Zeman1994} still recover isotropic turbulence properties.

The clear trend of a ``two-dimensionalization'' of the flow in model m15\_3D\_artrot\_2deg points towards an interesting feature of rotation in CCSN explosion models. 
Besides the additional amount of kinetic energy provided by rotation, also the turbulent energy cascade is (partially) reversed in the case of small Ro numbers. 
Similar to the behavior of 2D simulations \citep[cf.][]{Hanke2012}, fast rotation in 3D supports the development
of large-scale structures and the storage of energy on these scales, and therefore facilitates the evolution towards a successful explosion.
\newline

\subsection{A Rotation-supported Supernova Explosion in 3D} \label{sub:3D}

\begin{figure}
 \includegraphics[width=\columnwidth]{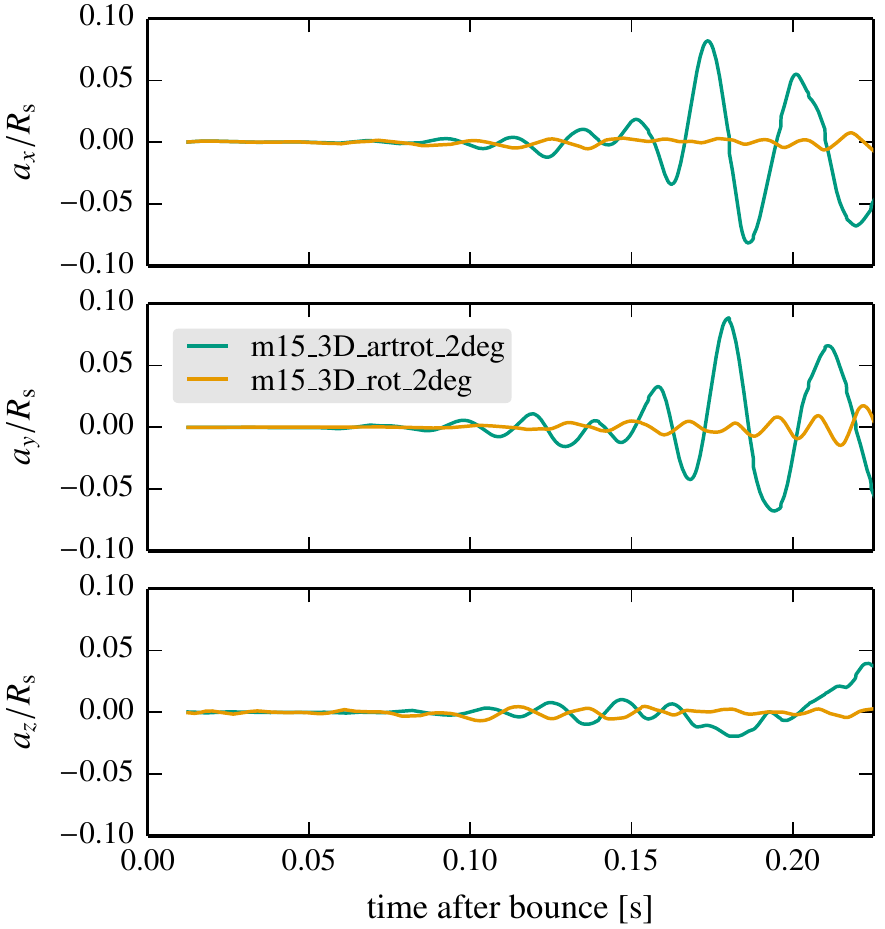}
 \caption{Time evolution of the coefficients $a_x$, $a_y$, and $a_z$ for an expansion of the shock surface into
 spherical harmonics (see definition in Sect.\ref{sub:3D}. The coefficients are normalized to the $l=0$, $m=0$ mode
 (i.e. the average shock radius).}\label{sasi_ampl}
\end{figure}

    \begin{figure*}
	    \subfigure{
            \includegraphics[width=0.32\textwidth]{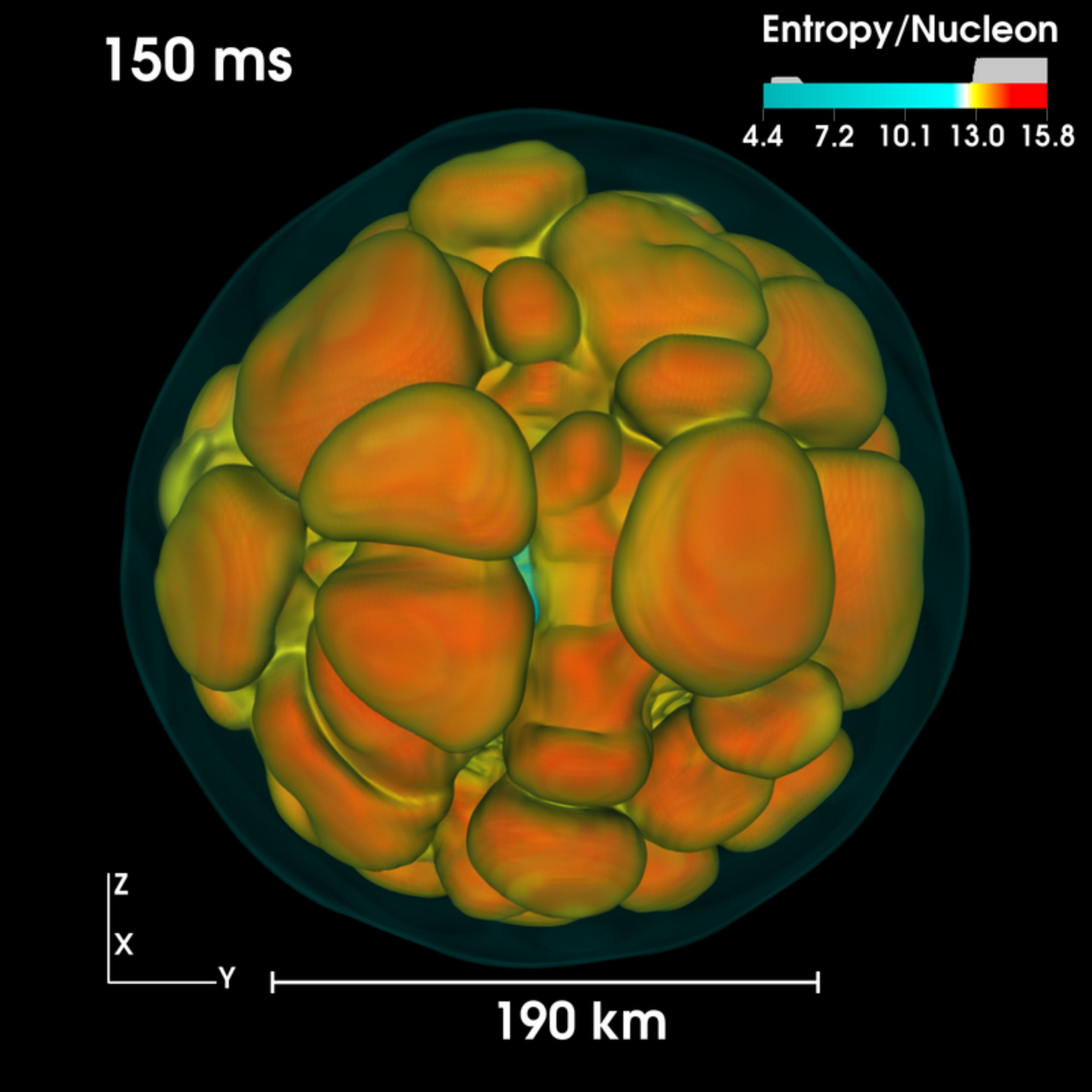}
}
        \hspace{-0.3cm}
	    \subfigure{  
            \includegraphics[width=0.32\textwidth]{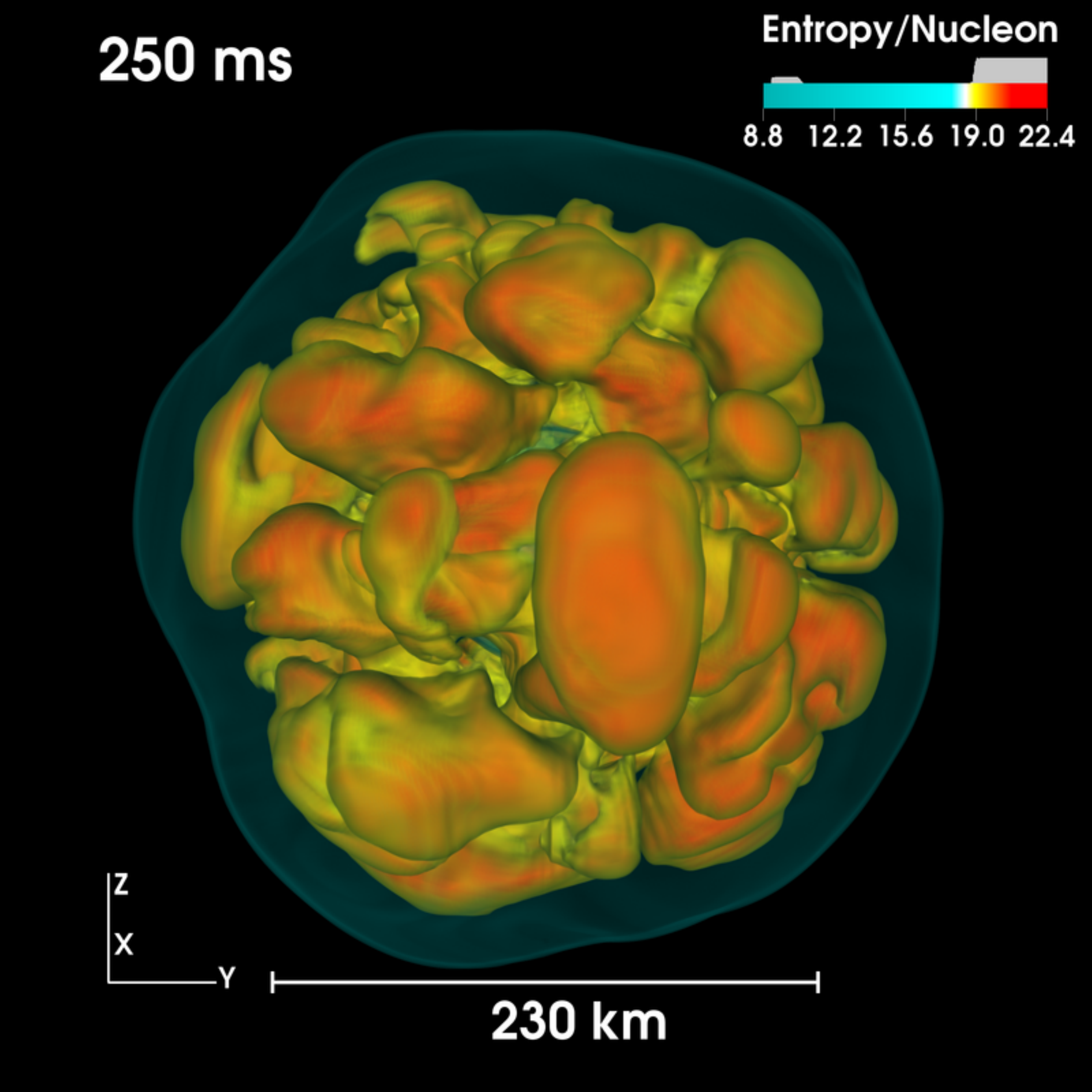}
}
        \hspace{-0.3cm}
	    \subfigure{  
            \includegraphics[width=0.32\textwidth]{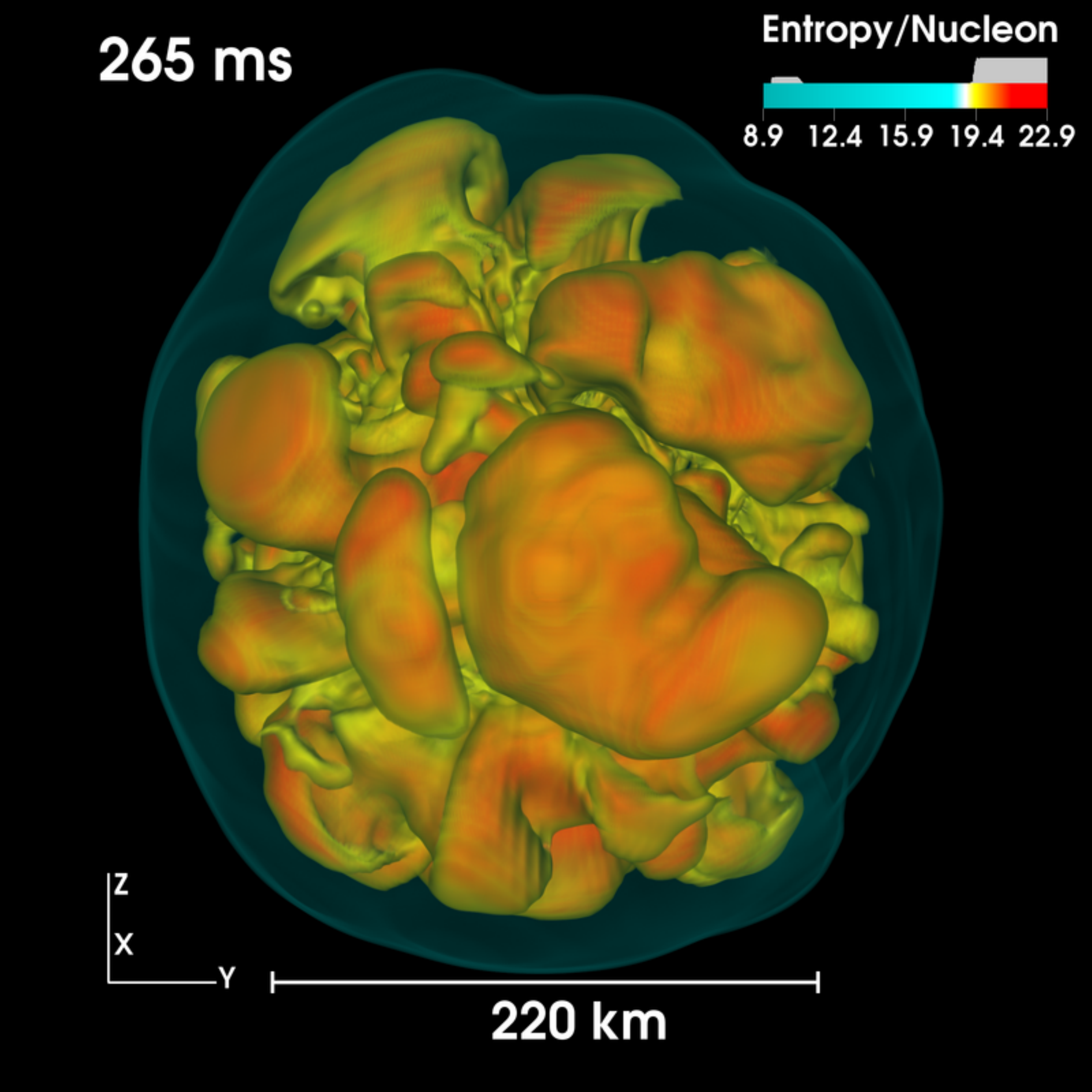}
}
        \vskip -0.25cm
	    \subfigure{
            \includegraphics[width=0.32\textwidth]{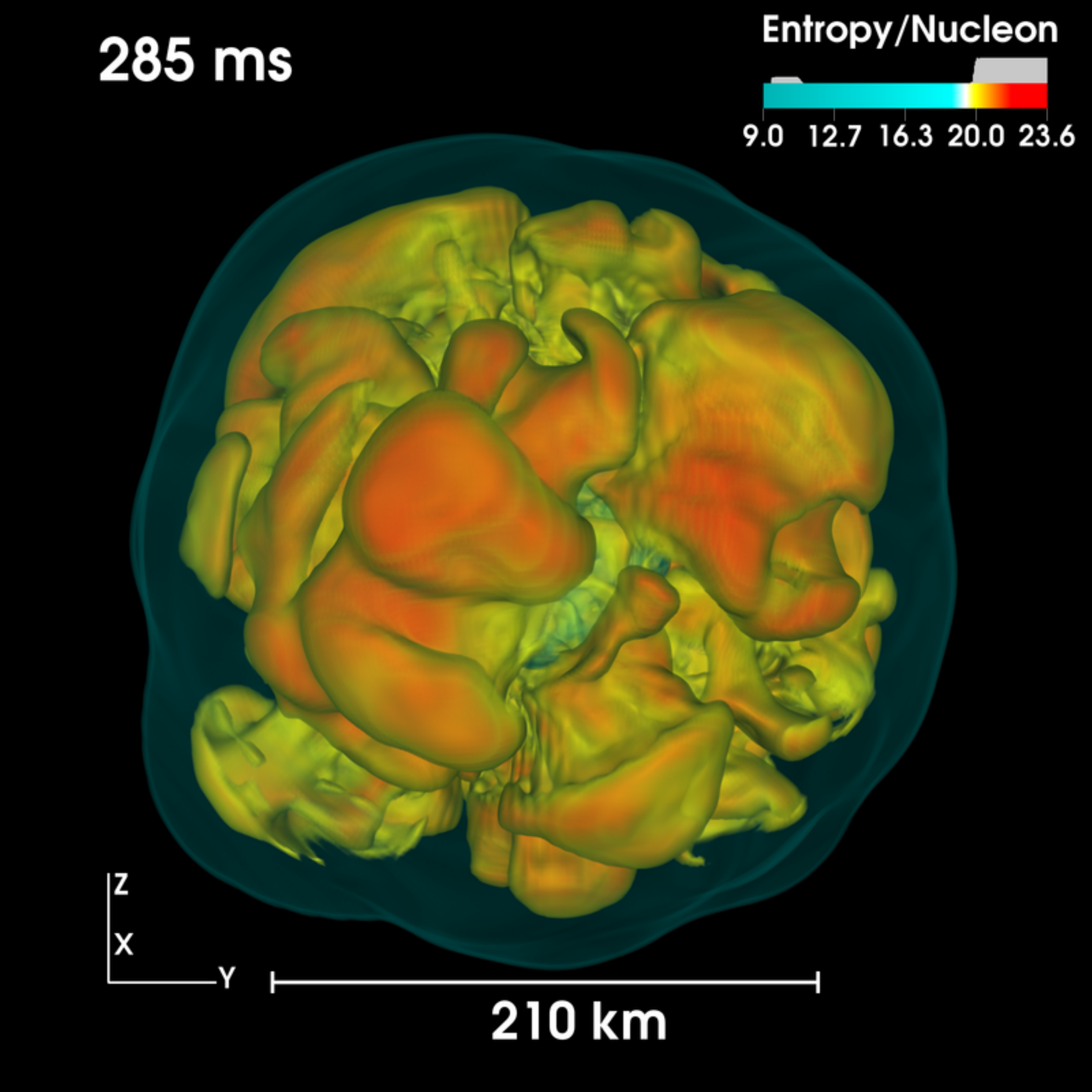}
}
        \hspace{-0.3cm}
	    \subfigure{  
            \includegraphics[width=0.32\textwidth]{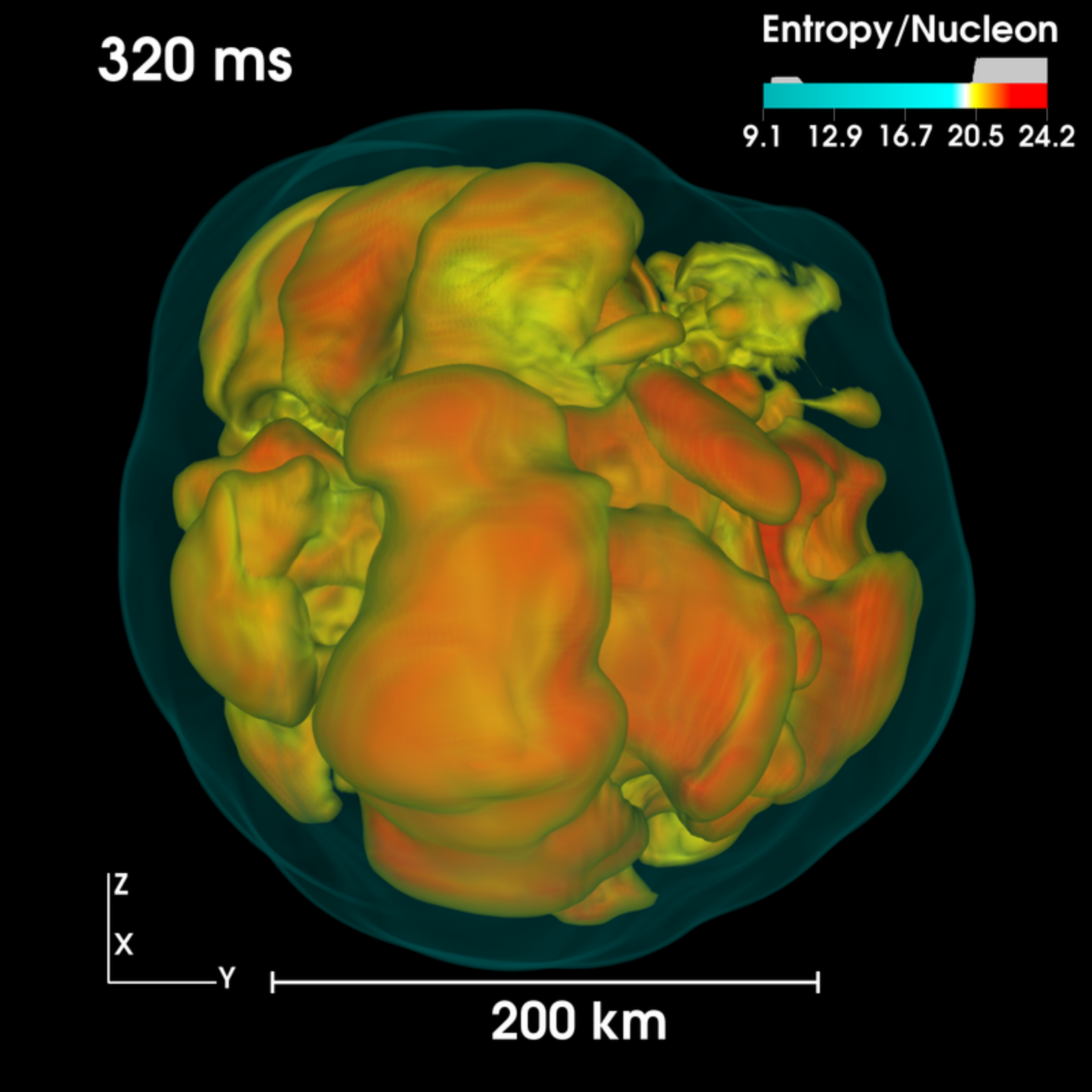}
}
        \hspace{-0.3cm}
	    \subfigure{  
            \includegraphics[width=0.32\textwidth]{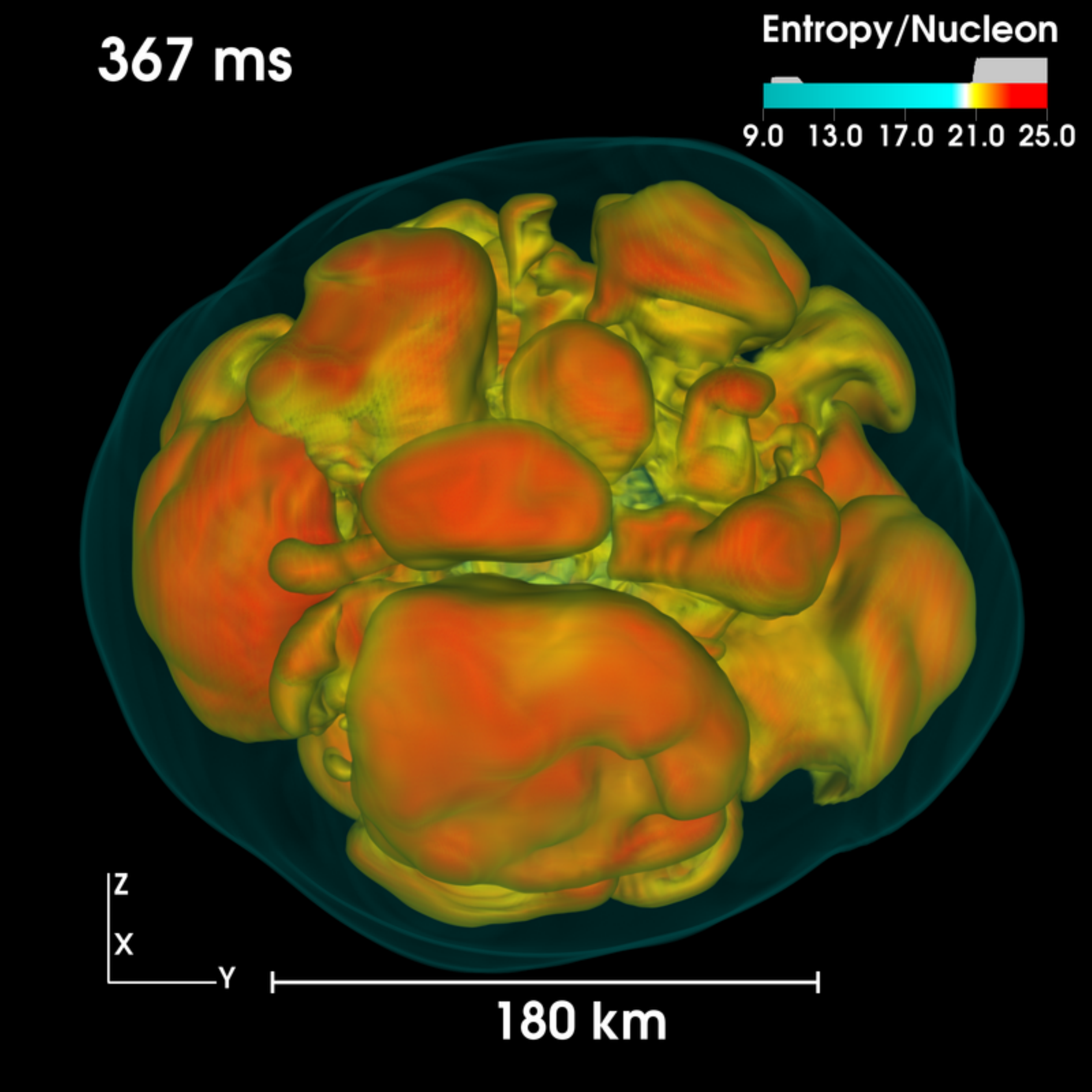}
}
        \caption{Volume rendered images of the nonexploding model m15\_3D\_rot\_2deg for different times
        after bounce. The rotation axis of the progenitor is in the $z$-direction. Colors represent entropy in $k_\mathrm{B}$ per nucleon, 
        the yardstick indicates the length scale. The supernova shock is visible by a thin bluish surface
        surrounding the high-entropy bubbles of neutrino-heated matter. The model is clearly convection
        dominated with only weak SASI activity.\label{m15_rot}}
    \end{figure*}

        \begin{figure*}
	    \subfigure{
            \includegraphics[width=0.32\textwidth]{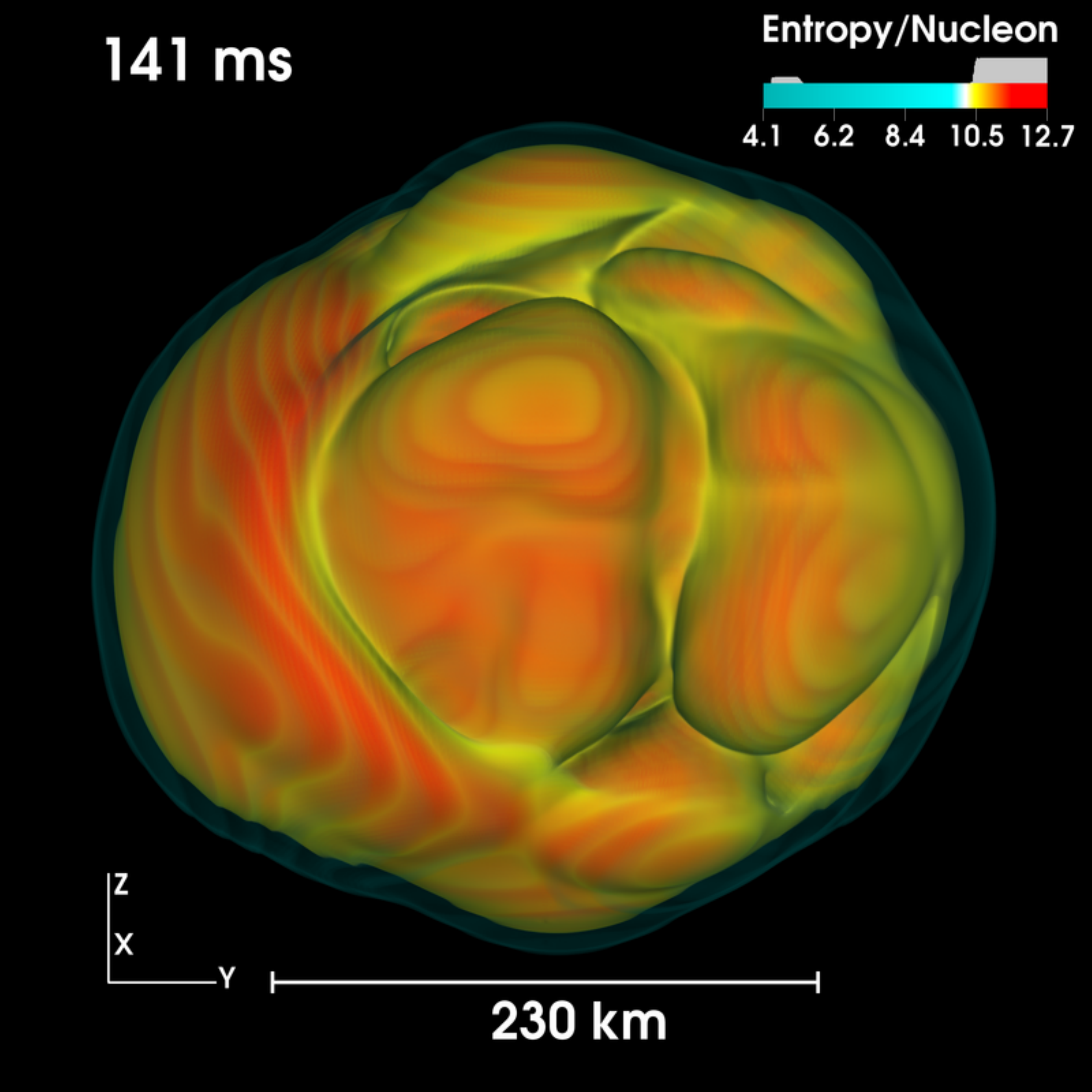}
}
        \hspace{-0.3cm}
	    \subfigure{  
            \includegraphics[width=0.32\textwidth]{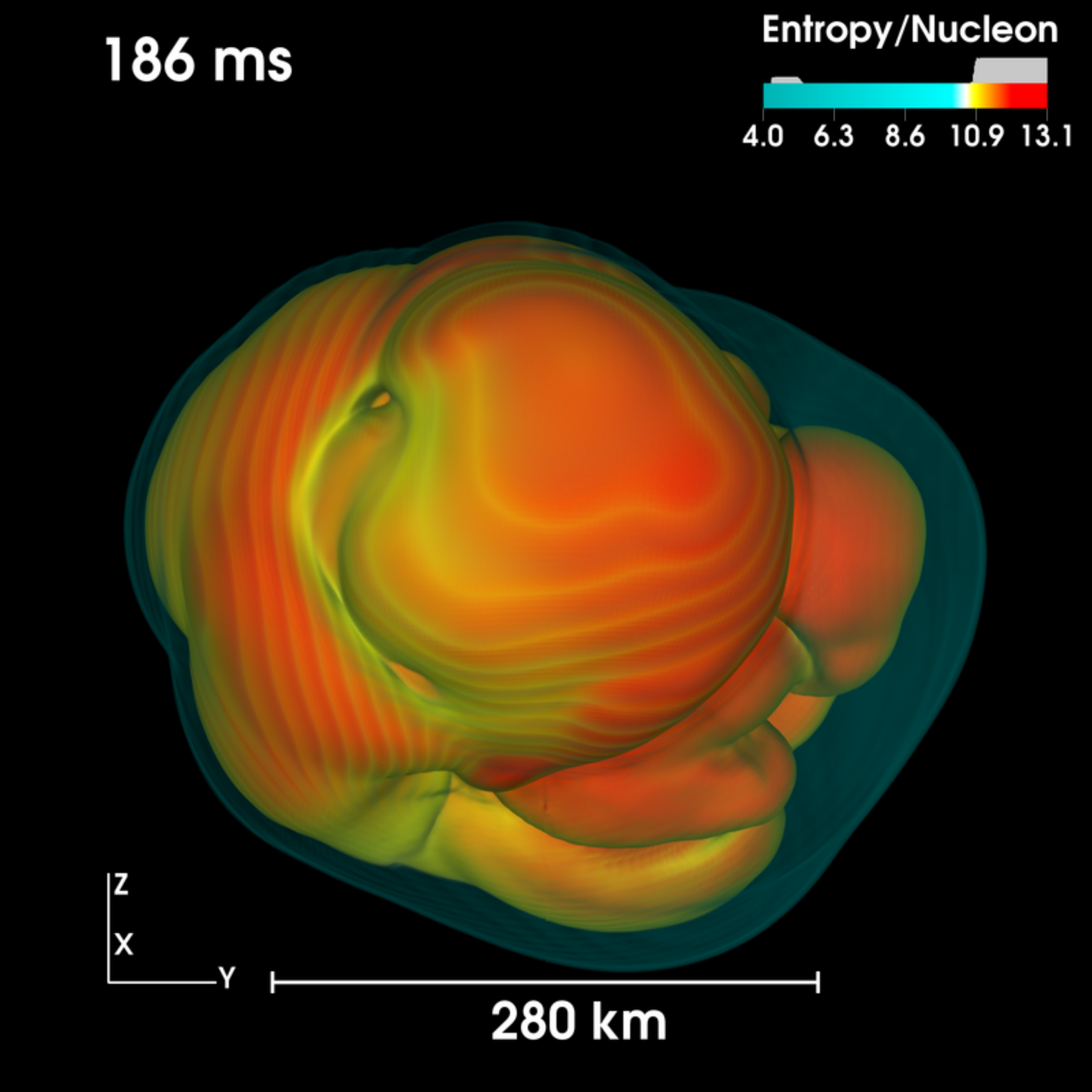}
}
        \hspace{-0.3cm}
	    \subfigure{  
            \includegraphics[width=0.32\textwidth]{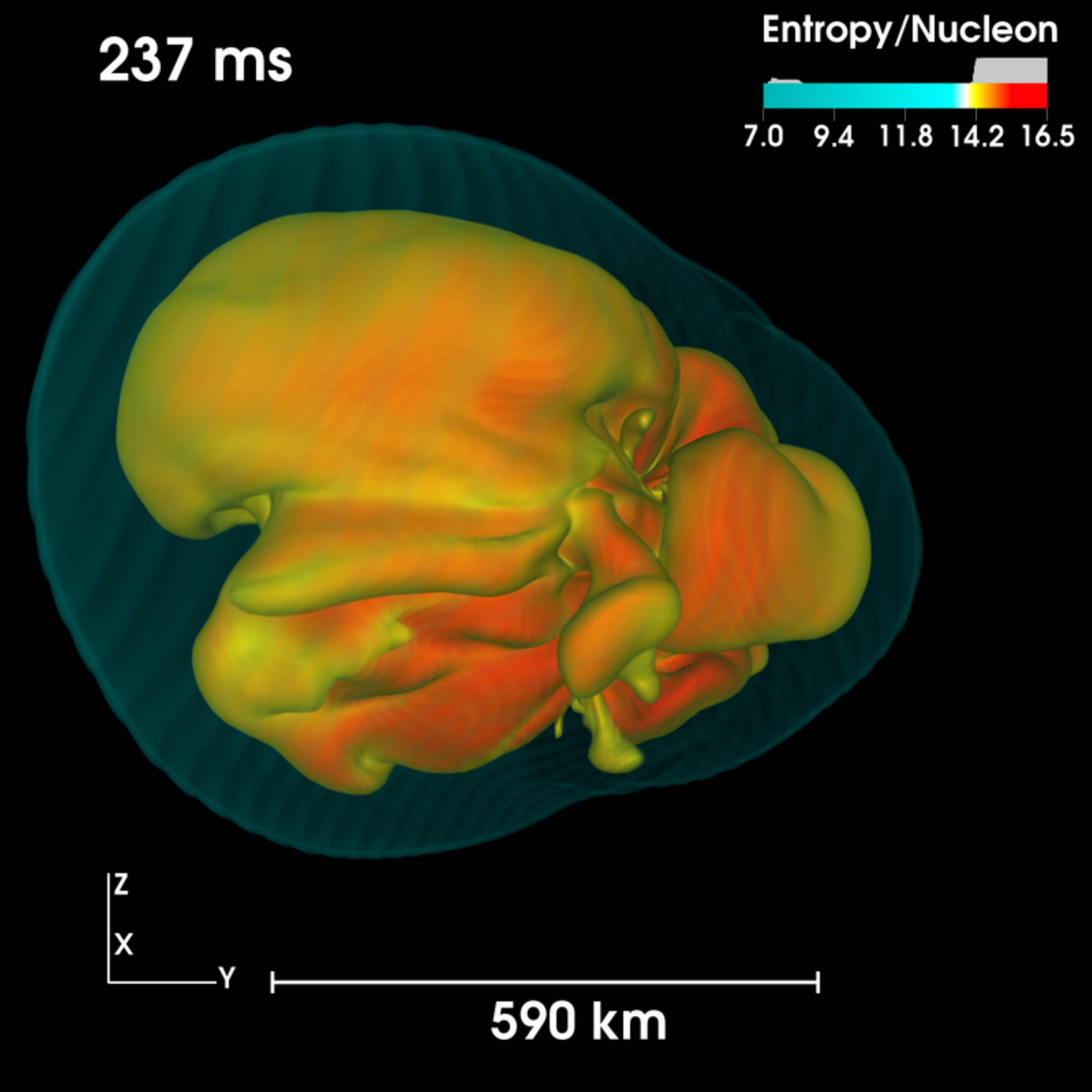}
}
        \vskip -0.25cm
	    \subfigure{
            \includegraphics[width=0.32\textwidth]{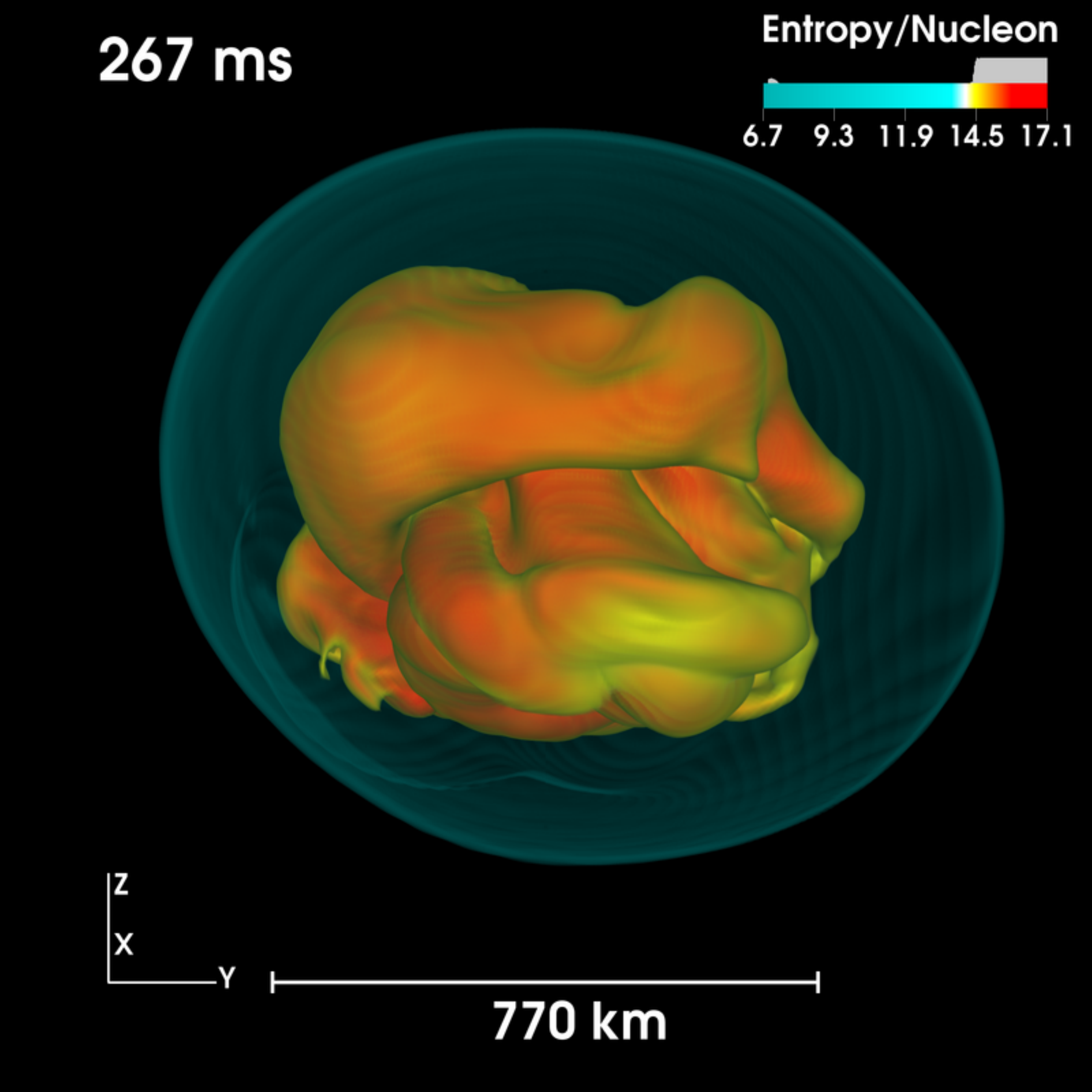}
}
        \hspace{-0.3cm}
	    \subfigure{  
            \includegraphics[width=0.32\textwidth]{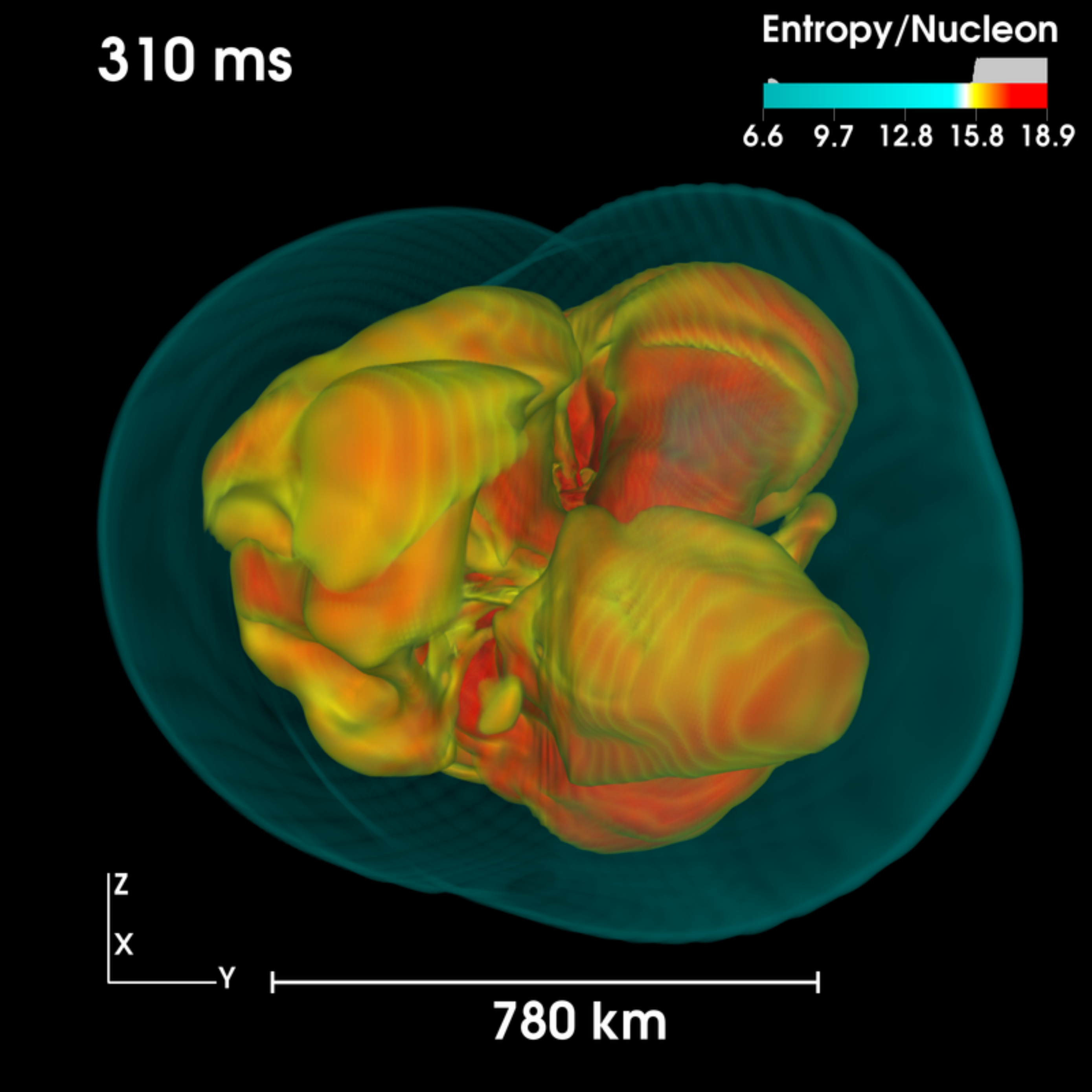}
}
        \hspace{-0.3cm}
	    \subfigure{  
            \includegraphics[width=0.32\textwidth]{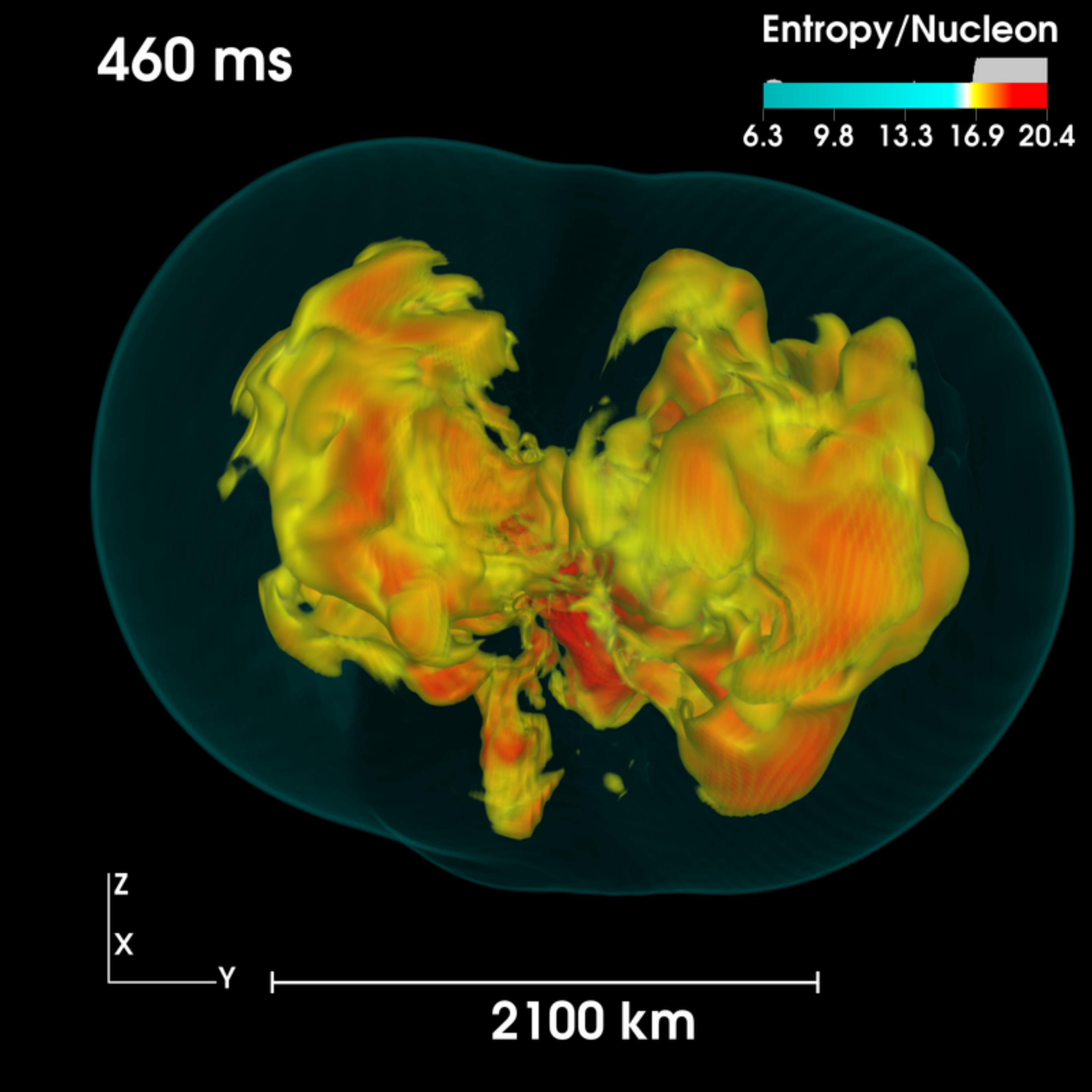}
}
        \caption{Volume rendered images of the exploding model m15\_3D\_artrot\_2deg for different times
        after bounce. The rotation axis of the progenitor is the $z$-direction. Colors represent entropy in $k_\mathrm{B}$ per nucleon, 
        the yardstick indicates the length scale. The supernova shock is visible by a thin bluish surface
        surrounding the high-entropy bubbles of neutrino-heated matter. The dynamics of the postshock layer
	are strongly dominated by a SASI spiral mode that supports an oblate-shaped explosion pushing the shock
	out in the equatorial plane perpendicular to the rotation axis of the progenitor model.\label{m15_artrot}} 
    \end{figure*}

The two best-resolved 3D models with moderate and enhanced rotation exhibit a largely different behavior:
The fast rotating model m15\_3D\_artrot\_2deg does not show any pronounced shock retraction phase 
and develops a strong spiral SASI mode giving support 
to a fairly early explosion even before the Si/Si-O interface reaches the shock (Figs.\,\ref{omega} and \ref{time_scales}).
In contrast, model m15\_3D\_rot\_2deg is far away from a successful explosion and the ratio of advection and heating time scale never 
increases beyond a value of 0.5 (see Fig.\,\ref{time_scales}). 

We emphasize once more that this behavior of the 3D models with rotation is in strong discrepancy to the results for the 
corresponding 2D simulations. In 2D, slow rotation might tend to support the explosion, and model m15\_2D\_rot\_2deg explodes somewhat
earlier than the nonrotating case m15\_2D\_norot\_2deg. However, our few models per case with a rather small difference 
in the explosion time are not sufficient to draw firm conclusions. Fast rotation in model m15\_2D\_artrot\_2deg, however,
delays the explosion (no success is obtained until the simulation was terminated at $\sim$\,470\,ms after bounce)
because of the reduced energy deposition by neutrinos in the gain layer, which is a consequence of lower radiated neutrino
luminosities and mean energies (Fig.\,\ref{nu_prop}). In the 3D model m15\_3D\_artrot\_2deg the rotationally induced
reduction of the emitted neutrino luminosities and mean energies is even more pronounced, but this effect is overcompensated
by the strong spiral SASI activity, which triggers the early explosion to set off in the equatorial plane. 

The oblate deformation of the beginning 3D explosion, with accretion continuing along the polar directions, agrees with the results for 
even more rapidly spinning progenitor conditions by \citet{Nakamura2014} and \citet{Takiwaki2016} and is also in stark
contrast to 2D explosions with and without rotation. In 2D, explosions start with a pronounced prolate deformation when
the outgoing shock expands fastest along the polar direction, driven either by the SASI sloshing or stronger neutrino heating
above the poles of the newly formed neutron star \citep[see Fig.\,\ref{entropy_cuts} and also][]{Kotake2006}.

In order to study the development of
the SASI activity in a quantitative way, we perform a decomposition of the angle-dependent shock radius 
$R_\mathrm{s}(\theta,\phi)$ into spherical harmonics $Y^m_l$:
\begin{equation}
 a^m_l = \frac{(-1)^{\left|m\right|}}{\sqrt{4\pi(2l+1)}}\int\limits_\Omega R_\mathrm{s}(\theta,\phi)\,Y^m_l(\theta,\phi)\,\mathrm{d}\Omega.
\end{equation}
The orthonormal basis functions are given by
\begin{equation}
 Y^m_l(\theta,\phi) = \sqrt{2} \,N^m_lP^m_l(\cos \theta) \cos(m\phi)
\end{equation}  
for $m > 0$, by
\begin{equation}
  Y^m_l(\theta,\phi) = N^0_lP^0_l(\cos \theta)
\end{equation}
for $m = 0$, and by
\begin{equation}
  Y^m_l(\theta,\phi) = \sqrt{2}\,N^{\left|m\right|}_lP^{\left|m\right|}_l(\cos \theta)\sin(\left|m\right|\phi)
\end{equation}
for $m < 0$, with
\begin{equation}
 N^m_l = \sqrt{\frac{(2l+1)(l-m)!}{4\pi(l+m)!}}
\end{equation}
and $P^m_l(\cos \theta)$ being the associated Legendre polynomials \citep{Burrows2012a,Ott2013}. 
With this choice of basis functions, the coefficients with $l=1$ represent the angle-averaged Cartesian coordinates of the shock surface, 
\begin{equation}
 a^{1}_1 = \langle x_\mathrm{s} \rangle \equiv a_x,
\end{equation}
\begin{equation}
 a^{-1}_1 = \langle y_\mathrm{s} \rangle \equiv a_y,
\end{equation}
\begin{equation}
 a^{0}_1 = \langle z_\mathrm{s} \rangle \equiv a_z.
\end{equation}

The time evolution of the SASI amplitudes can be inferred from Fig.\,\ref{sasi_ampl}. Although weak oscillations
of the shock surface also arise in the case of model m15\_3D\_rot\_2deg at 100\,ms after bounce, a further growth
of the amplitudes seems to be suppressed and the fluid motions in the gain layer are dominated by convection 
(see also Fig.\,\ref{m15_rot}). While the advection time scales in both models are similar only 
at very early post-bounce times (cf.\ Fig.\,\ref{time_scales}, second panel from top), model m15\_3D\_artrot\_2deg first shows
influence by the accumulating rotational energy in the gain layer and then exhibits a strong growth
of a spiral SASI mode in the $x$-$y$-plane, which is perpendicular to the rotation axis of the infalling matter.
The SASI motions drive the shock to larger and larger 
radii in the equatorial region and develop an oblate ejecta deformation (see Fig.\,\ref{m15_artrot}).
At later times, we observe the presence of a corotation radius \citep[cf.][]{Kazeroni2017}, which means there is a location
between neutron star and shock where the flow rotates faster than the SASI spiral pattern of the shock. Since the initial angular 
velocities are not as extreme as the rapid rotation rates assumed by \citet{Takiwaki2016}, no one-armed spiral
wave pattern emerges. Model m15\_3D\_artrot\_2deg resembles an intermediate case situated between the two models presented in
figure 4 of \citet{Kazeroni2017}: The initially SASI-dominated rotation dynamics transitions to a phase where the innermost part
of the gain region rotates faster than the spiral SASI pattern, even though a tightly-wound spiral flow being characteristic for 
a pronounced corotational instability does not appear.

Due to the tremendous computational costs for 3D simulations with state-of-the-art neutrino transport,
we cannot investigate the impact of rotation on the development of SASI for an extended set of models similar to the
study by \citet{Kazeroni2017}, who used largely simplified simulation setups. Especially in view of the stochastic nature of the SASI 
\citep[cf.][]{Kazeroni2017}, the limited 3D model set presented in this paper does not allow any comprehensive
evaluation of the impact of rotation on the growth of the SASI. But the fact that, on the one hand, no spiral SASI mode
is triggered in the slowly rotating model m15\_3D\_rot\_2deg and, on the other hand, a \textit{sloshing} SASI mode emerges 
in its less-resolved counterpart m15\_3D\_rot\_6deg (see the large excursions of the average shock radius of this model in Fig.\,\ref{shock_radii}), 
strongly suggests that the rotation rates obtained in current stellar evolution calculations including magnetic fields 
are not high enough to modify the explosion physics in a significant way compared to nonrotating progenitors. 

\begin{figure}
 \includegraphics[width=\columnwidth]{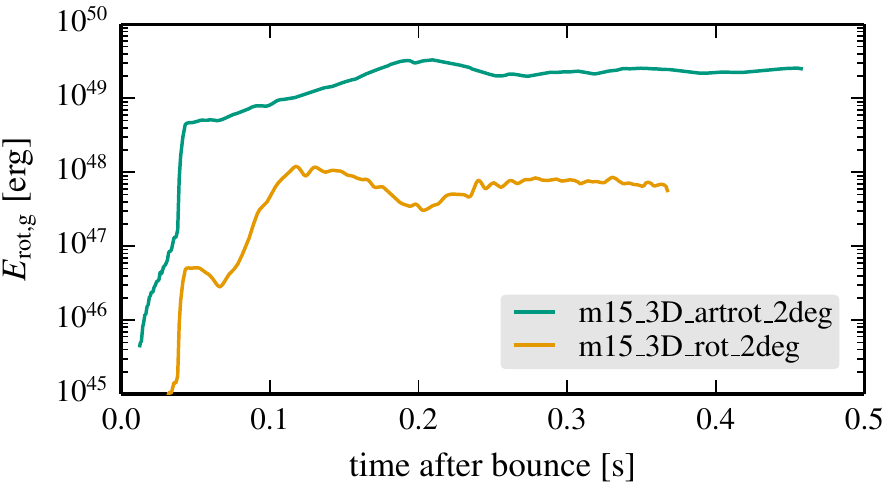}
 \caption{Time evolution of the rotational energy in the gain layer for the two best-resolved 3D models. 
 The curves are smoothed by running averages of 5\,ms.\label{e_rot}}
\end{figure}

This conclusion is further supported by a comparison of 
the rotational energy contained by the gain layer in models m15\_3D\_rot\_2deg and m15\_3D\_artrot\_2deg (Fig.\,\ref{e_rot}). This quantity
is at least one order of magnitude higher in model m15\_3D\_artrot\_2deg during the linear growth phase of SASI spiral motions
in the fast rotating case around
100\,ms after bounce, underlining that the rotation velocities of the accreted material in model m15\_3D\_rot\_2deg
are too low to foster the creation of a spiral SASI mode. For such slow rotation with Rossby numbers $\mathrm{Ro}\gg1$, centrifugal forces play a 
negligible role and the critical luminosity to be reached for a successful shock runaway is hardly reduced 
compared to the nonrotating case (see also Sect.\,\ref{sec:crit}). 

\begin{figure}
 \includegraphics[width=\columnwidth]{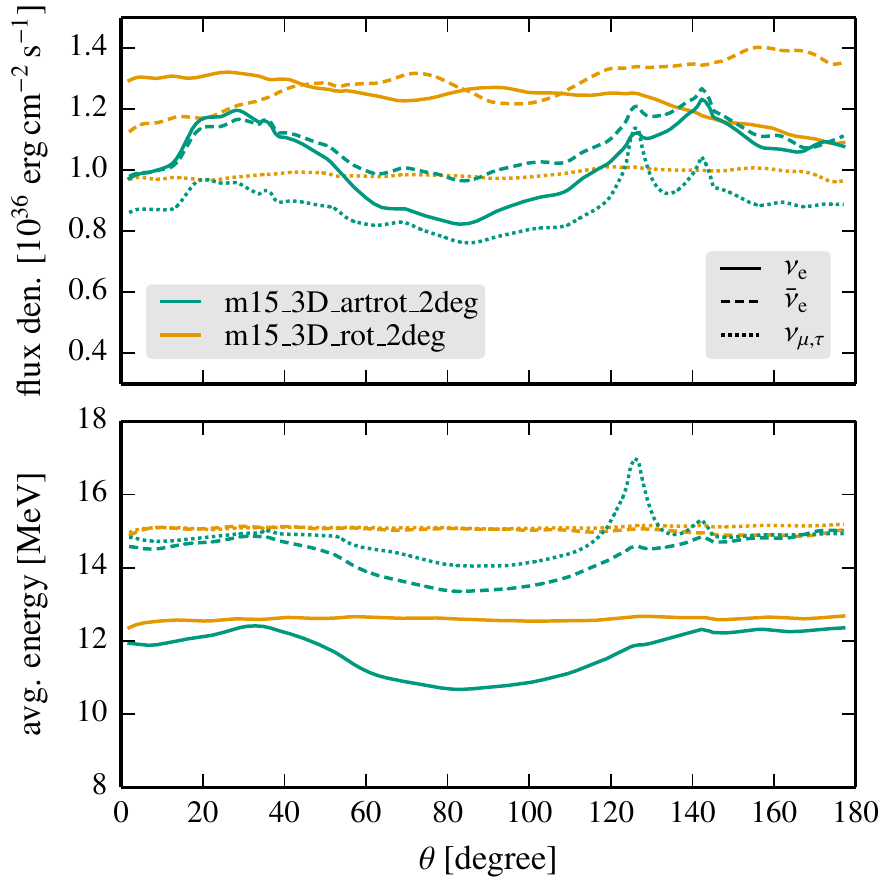}
 \caption{Neutrino energy flux densities and mean energies as functions of the polar angle at 365\,ms after bounce. The lab-frame quantities are 
 measured at a radius of 400\,km and averaged over the azimuthal angle.}\label{fnu}
\end{figure}

The stronger influence of rotation in model m15\_3D\_artrot\_2deg cannot only be seen by the violent spiral SASI
activity and a considerable overall reduction of the radiated neutrino luminosities and mean energies (Fig.\,\ref{nu_prop}),
but also by a characteristic pole-to-equator variation of the neutrino emission.
Rotation leads to a deformation of the proto-neutron star and predominant accretion by the neutron star from 
the polar directions at later times, which leaves its imprint on the 
angle-dependent neutrino emission (see Fig.\,\ref{fnu}). In the depiction of the 
neutrino energy flux densities and mean energies versus polar angle, the fast rotating model exhibits a clear trough-like shape
\citep[see the detailed discussion of similar 2D results in section 3.6 of][]{Marek2009a}.
Such an effect is not visible for model m15\_3D\_rot\_2deg.

In analogy to the rotating 2D model discussed by \citet{Marek2009a}, also in the 3D situation of model m15\_3D\_artrot\_2deg,
the polar accretion flows produce emission maxima for all neutrino species at angles between 20 and 40 degrees away from 
the pole, but these peaks are somewhat less sharp and lower than the emission maxima in the 2D case of \citet{Marek2009a}.
This difference might be a consequence of the unintended loss of angular momentum during the collapse phase in our
present models, which reduces the rotational deformation of the newly formed neutron star compared to the previous publication.

In Fig.\,\ref{time_scales} (bottom panel), the diagnostic explosion energy versus time is plotted.
It is given by 
\begin{equation}
E_\mathrm{diag}=\int\limits_{V_g} 
e_{\mathrm{tot,>0}}\,\rho\,\mathrm{d}V,
\end{equation}
which is calculated as the integral of the total 
(i.e.\ kinetic plus internal plus gravitational) specific energy, defined as
$e_\mathrm{tot}=v^2/2+\epsilon+\Phi$, over the gain layer, including all regions where
this energy is positive. Following \citet{Melson2015}, we assume that all nucleons
finally recombine to iron-group nuclei. At the time the simulation had to be stopped 
because of the extremely high computational demands, model m15\_3D\_artrot\_2deg 
reaches a diagnostic energy of $\sim9.3\times 10^{49}\,\mathrm{erg}$, which is 
still rising steeply. Owing to the oblate explosion geometry (see Figs.\,\ref{entropy_cuts}
and \ref{m15_artrot}), the ongoing accretion in the polar regions is not expected 
to cease soon. We therefore expect a considerable increase of the diagnostic energy
even at later times after the onset of shock expansion as long as simultaneous accretion and shock
expansion take place, an effect that was predicted by \citet{Marek2009a} and is seen in long-time 2D
simulations \citep{Bruenn2016,Nakamura2016} as well as recent 3D simulations of nonrotating
progenitors \citep{Mueller2015a,Mueller2017}.

\subsection{Resolution Dependence}
Although we have briefly discussed resolution-dependent aspects of our results
in the previous sections, we summarize the main points in the following. In 2D,
the models with an angular resolution of 1.4 and 2 degrees evolve quite similarly
(see Fig.\,\ref{shock_radii}, lower panel). While the better resolved simulations
seem to exhibit a weak tendency to explode slightly earlier (Table~\ref{tab:table}), 
the observed differences between the two simulations for each rotation
profile are in the ballpark of stochastic fluctuations \citep[see][]{Summa2016}.
In 3D, the setup chosen for the low-resolution runs with only 4 or 6 degrees in 
the angular domain (but the same radial resolution)
was a compromise between the computational costs and our curiosity
to investigate a larger set of models without real confidence in the numerical
convergence of some of them. The coarseness of the grid for the less resolved models
clearly has an impact on the temporal evolution. In the cases where a comparison is
possible we find that the low-resolution models show a transient shock expansion
up to 200\,km after the initial shock retraction phase, in contrast to a better resolved
model with an angular resolution of 2 degrees. The wider excursions of the average shock
radius are a reaction to enhanced SASI activity during the corresponding time interval.
We interpret this difference as a consequence of the suppressed growth of parasitic 
Rayleigh-Taylor and Kelvin-Helmholtz instabilities in models with lower angular resolution.
In models with higher angular resolution these instabilities allow energy to cascade
by turbulent fragmentation of vortex flows from the large SASI scales to smaller scales,
thus damping the SASI amplitudes \citep{Guilet2010}. For this reason our low-resolution
results have to be taken with caution and the major part of this paper focuses on
the best-resolved 3D runs in our set with an angular resolution of 2 degrees.

\section{The Critical Neutrino Luminosity Condition Revisited} \label{sec:crit}

\begin{figure*}
\centering
\includegraphics[width=0.83\textwidth]{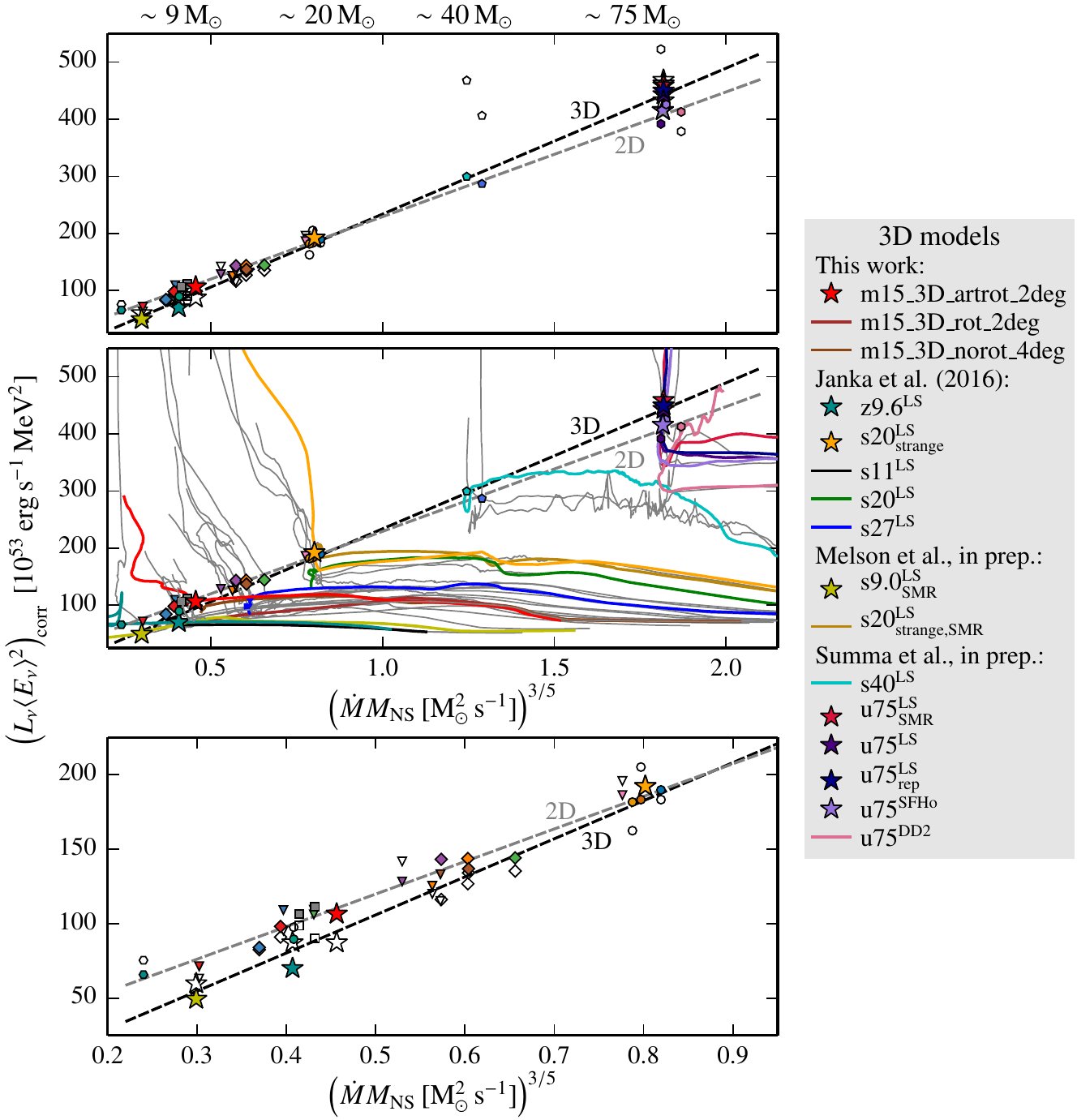}
\caption{Critical luminosity relations for explosion or the onset of strong shock
expansion, $(L_\nu\langle E_\nu^2\rangle)_\mathrm{crit,corr} \propto
(\dot{M}M_\mathrm{NS})^{3/5}$ of Eq.~(\ref{eq:crit_cond}), for our
considered sets of 2D (gray dashed line) and 3D simulations
(black dashed line). The bottom panel provides a zoom to the 
progenitor-mass region of $\lesssim$\,20\,M$_\odot$.
The lines are obtained by least-squares fits to the filled
symbols for ``corrected'' values of the heating functional (Eq.~\ref{eq:crit_cond})
at the critical locations along the evolution tracks that are displayed by curves in
the middle panel, considering 2D and 3D model sets separately.
The best-fit lines of the available simulations show a modest dependence on dimension.
The open symbols denote the
uncorrected values of $(L_\nu\langle E_\nu^2\rangle)_\mathrm{crit}$.
All symbols mark points on the evolution tracks when the time-scale
ratio $\tau_\mathrm{adv}/\tau_\mathrm{heat}$ reaches a value of 1.2. Big colored
asterisks and colored evolution tracks denote 3D models, whereas the small symbols
and grey curves in the middle panel belong to 2D simulations. 
The list to the right of the panels provides an 
overview of all included 3D simulations (see text in Sect.\,\ref{model_set}). 
Cases with explosions (progenitors below
$\sim$20\,M$_\odot$, see approximate mass scale above the top panel) 
or strong shock expansion (black-hole forming progenitors of 40--75\,M$_\odot$)
have a line with an asterisk, the others just a line. The symbols and the color coding 
of the 2D models are adopted from figure 14 of \citet{Summa2016} and figure 3 
of \citet{Janka2016} for previously published results \citep[for details, see the
list of models in][]{Summa2016}. Symbols for black-hole forming
2D models possess the same colors as the corresponding 3D cases. We emphasize
that the progenitor-mass scale on top of the figure is supposed to reflect only a 
rough, overall trend, in which lower-mass stars tend to be located towards the left
while higher-mass stars tend to lie towards the right. However, local inversions
with progenitor mass can occur because $\dot M$ and $M_\mathrm{NS}$ correlate with
the compactness of the stellar core \citep{Nakamura2015}, but the compactness is 
not a monotonic function of the progenitor mass \citep{OConnor2011}.
\label{crit_lum}}
\end{figure*}

Based on a toy model setup for steady-state accretion including neutrino heating behind the stalled
accretion shock, \citet{Burrows1993} showed that for a given mass-accretion rate $\dot{M}$, there exists a critical 
neutrino luminosity $L_{\nu,\mathrm{crit}}(\dot{M})$ above which a stationary accretion flow onto the 
proto-neutron star is no longer possible. \citet{Janka2012a} demonstrated by analytic arguments that this critical
luminosity condition is equivalent to the condition of $\tau_\mathrm{adv}/\tau_\mathrm{heat}\gtrsim1$ for 
the ratio of advection and heating time scales \citep[for hydrodynamical toy-model simulations in 1D leading to
the same conclusion, see][]{Fernandez2012}.

\subsection{Generalized Critical Heating Condition}

In \citet{Mueller2015} this concept was generalized to a critical condition for the `heating functional'
$(L_\nu\langle E_{\nu}^2\rangle)_\mathrm{crit}$ in dependence on the mass-accretion rate $\dot{M} $, the proto-neutron star mass $M_\mathrm{NS}$
and the gain radius $R_\mathrm{g}$. Because we compare results for different progenitors in the present
work \citep[in contrast to][]{Mueller2015}, we use the version of the critical condition
introduced by \citet{Summa2016}, equation 30, in which all gain-layer dependent properties are subsumed in a factor $\xi_\mathrm{g}$:
\begin{equation}
 \left(L_\nu\langle E_{\nu}^2\rangle\right)_\mathrm{crit} \propto \left(\dot{M}M_\mathrm{NS}\right)^{3/5} \xi_\mathrm{g}.
  \label{eq:nu_crit}
\end{equation}
With this definition, the correction factor $\xi_\mathrm{g}$ accounts for 
progenitor-dependent variations of the gain radius and 
turbulent effects that directly influence the critical neutrino luminosity.

\citet{Summa2016} and \citet{Janka2016} further generalized the critical condition to include the effects of rotation
and to take care of case-dependent differences of the total energy at the gain radius. This was achieved by
modifications of the correction factor $\xi_\mathrm{g}$, whose exact functional form can be redefined to also account for
the effects on the shock evolution associated with large-scale velocity perturbations in the convective Si and O burning
shells prior to core collapse \citep{Mueller2016a,Mueller2017}.

The main steps in deriving this critical luminosity can be summarized as follows.
By the use of approximate scaling relations for the advection timescale $\tau_\mathrm{adv}$,
\begin{equation}
\tau_\mathrm{adv} \propto \frac{R_\mathrm{s}^{3/2}}{\sqrt{M_\mathrm{NS}}},
\label{eq:t_adv}
\end{equation}
the heating timescale $\tau_\mathrm{heat}$,
\begin{equation}
\tau_\mathrm{heat} \propto \frac{\left|\bar{e}_\mathrm{tot,g}\right|R_\mathrm{g}^2}{L_\nu \langle E_\nu^2\rangle},
\end{equation}
and the radius of the stalled shock in spherical symmetry,
\begin{equation}
R_\mathrm{s} \propto \frac{\left(L_\nu \langle E_\nu^2\rangle\right)^{4/9}R_\mathrm{g}^{16/9}}{\dot{M}^{2/3}M_\mathrm{NS}^{1/3}}
\label{eq:r_shock}
\end{equation}
\citep[see][]{Janka2012a,Mueller2015}, the timescale criterion $\tau_\mathrm{adv}/\tau_\mathrm{heat}\sim1$ as conventional condition for
runaway shock expansion \citep[e.g.][]{Thompson2005,Buras2006a,Marek2009a,Fernandez2012} can directly be translated into a condition for the critical luminosity or heating functional:
\begin{equation}
\left(L_\nu \langle E_\nu^2\rangle\right)_\mathrm{crit}\propto\left(\dot{M}M_\mathrm{NS}\right)^{3/5}\left|\bar{e}_\mathrm{tot,g}\right|^{3/5}R_\mathrm{g}^{-2/5}.
\end{equation}
Here, $L_\nu$ denotes the total luminosity of electron neutrinos and antineutrinos (because both of these neutrino species dominate the neutrino heating), 
$L_\nu=L_{\nu_\mathrm{e}}+L_{\bar{\nu}_\mathrm{e}}$, and $\langle E_\nu^2\rangle$ is 
defined as the weighted average of the corresponding mean squared energies,
\begin{equation}
\langle E_\nu^2\rangle = \frac{L_{\nu_\mathrm{e}} \langle E_{\nu_\mathrm{e}}^2\rangle + L_{\bar{\nu}_\mathrm{e}} \langle E_{\bar{\nu}_\mathrm{e}}^2\rangle}{L_{\nu_\mathrm{e}}+L_{\bar{\nu}_\mathrm{e}}}.
\end{equation}
The mean squared energies stand for the squared energies of the neutrino-energy distributions, which can be expressed 
in terms of the energy moments of the neutrino-number distributions, 
$\langle E_{\nu_\mathrm{e}}^2\rangle \coloneqq \langle \epsilon_{\nu_\mathrm{e}}^3\rangle / \langle \epsilon_{\nu_\mathrm{e}}\rangle$ 
and $\langle E_{\bar{\nu}_\mathrm{e}}^2\rangle \coloneqq \langle \epsilon_{\bar{\nu}_\mathrm{e}}^3\rangle / \langle \epsilon_{\bar{\nu}_\mathrm{e}}\rangle$.

In addition to the dependence on the gain radius $R_\mathrm{g}$ and on the average mass-specific binding energy in the gain layer,
\begin{equation}
\bar{e}_\mathrm{tot,g} = \frac{E_\mathrm{tot,g}}{M_\mathrm{g}},
\end{equation}
the time dependent factor $\xi_\mathrm{g}$ in Eq.\,(\ref{eq:nu_crit}) can absorb further corrections
that are necessary to generalize the critical luminosity condition to the multidimensional case:
\begin{equation}
 \xi_\mathrm{g} \equiv \left|\bar{e}_\mathrm{tot,g}\right|^{3/5} R_\mathrm{g}^{-2/5} \xi_\mathrm{turb}^{-3/5} \xi_\mathrm{rot}^{6/5}
 \label{eq:corr}
\end{equation}
\citep{Janka2016}.

The consideration of nonradial (turbulent) postshock flows through an additional isotropic pressure contribution
$P_\mathrm{turb} \approx \langle v_\mathrm{aniso}^2\rangle\rho \approx 4/3 \langle\mathrm{Ma}^2\rangle P$ leads to the
definition of $\xi_\mathrm{turb}$ \citep[cf.][]{Mueller2015},
\begin{equation}
 \xi_\mathrm{turb} \equiv 1 + \frac{4}{3}\langle\mathrm{Ma}^2\rangle,
 \label{eq:xi_turb}
\end{equation}
where we follow \citet{Janka2016} by using the definition
\begin{equation}
\begin{split}
 \langle \mathrm{Ma}^2 \rangle & = \frac{\langle v_\mathrm{aniso}^2\rangle}{\langle c_\mathrm{s,g}^2\rangle} = \frac{1}{\langle c_\mathrm{s,g}^2\rangle}\sum\limits_{i=r,\theta,\phi}\langle\left(v_i - \bar{v}_i\right)^2\rangle \\ 
& = \frac{2E_\mathrm{kin,g}^\mathrm{aniso}/M_\mathrm{g}}{\langle c_\mathrm{s,g}^2\rangle}.
\end{split}
\end{equation}
$\bar{v}_{r,\theta,\phi}$ are angular averages over spherical shells, i.e.\ ordered radial flows due to accretion or the
expansion of the gain layer and coherent angular motions caused by stellar rotation or spiral SASI modes are subtracted.
The sound speed $c_\mathrm{s,g}$ is directly extracted from the numerical simulations as a mass-weighted average over
the gain layer:
\begin{equation}
\langle c_\mathrm{s,g}^2 \rangle = \frac{1}{M_\mathrm{g}}\int\limits_{V_\mathrm{g}} c_\mathrm{s}^2 \rho\, \mathrm{d}V.
\end{equation}

Note that in contrast to \citet{Mueller2015} and \citet{Mueller2016,Mueller2017} we prefer to evaluate $\langle c_\mathrm{s,g}^2 \rangle$
as numerical average over the gain layer instead of estimating a postshock value from the shock-jump conditions. 
The volume-integrated approach makes the analysis numerically more robust, but might not be fully compatible with the fact that
local properties at the gain radius or behind the shock could more directly determine the dynamical behavior.

The effects of rotation are included by an additional correction factor
\begin{equation}
 \xi_\mathrm{rot} \equiv \sqrt{1-\frac{j_0^2}{2GM_\mathrm{NS}R_\mathrm{s}}} \leq 1,
\end{equation}
where $j_0$ is defined as the average specific momentum on spherical shells, which is essentially conserved during the
infall of the matter in the pre-collapse region. The factor $\xi_\mathrm{rot}$ accounts for the reduction of the infall
velocity ahead of the stalled shock due to centrifugal effects, which leads to an increase of the accretion time scale in the 
postshock layer (Eq.\,\ref{eq:t_adv}) as well as to an increase of the shock-stagnation radius (Eq.\,\ref{eq:r_shock}).
For a detailed discussion, see \citet{Janka2016}.

As also detailed in \citet{Janka2016}, rotation therefore decreases the critical luminosity necessary for a runaway shock expansion or a successful explosion. 
Besides the effects of rotation taken into account by the correction factor $\xi_\mathrm{rot}$ mentioned above, rotational energy in
the gain layer also shifts the (negative) total specific energy $\bar{e}_\mathrm{tot,g}$ closer to zero and leads to an additional
reduction of the critical luminosity (see Eqs.\,\ref{eq:nu_crit} and \ref{eq:corr}).

\subsection{Critical Luminosity Relation for a Large Set of Models}\label{model_set}

With a growing set of multidimensional simulations becoming available, we can test the validity of the critical condition 
(Eq.\,\ref{eq:nu_crit} with Eq.\,\ref{eq:corr}) for an increasingly wider range of progenitor conditions. While \citet{Summa2016}
already used a larger sample of 2D models, we include here our 2D and 3D cases with rotation. Moreover, we also take into
account 3D results of nonrotating 9 and 20\,M$_\odot$ stars as well as  calculations for black-hole forming 40 and 75\,M$_\odot$
progenitors \citep{Woosley2002,Woosley2007}, all of which will be published in detail in forthcoming papers 
(Melson et al., in preparation, and Summa et al., in preparation, respectively). 
In the model names, the superscripts denote the employed high-density EoSs. LS indicates the 
EoS of \citet{Lattimer1991} with a nuclear incompressibility of 220\,MeV, SFHo the EoS introduced by 
\citet{Hempel2012} and \citet{Steiner2013}, 
and DD2 the EoS provided by \citet{Typel2010} and \citet{Fischer2014}. The subscript ``SMR'' refers to the 
application of a static angular mesh refinement technique for the computational grid 
\citep{Melson2016}, and the subscript ``strange'' denotes the use
of a reduced neutral-current neutrino-nucleon scattering opacity motivated by possible 
strangeness contributions to the nucleon spin, 
which affect the axial-vector weak coupling \citep{Melson2015a}. In the case of model 
u75$^\mathrm{LS}_\mathrm{rep}$, the subscript
just indicates a repetition of model u75$^\mathrm{LS}$ with a parallelization over the energy bins, 
which was newly implemented in our neutrino-transport code.

Including these still unpublished results allows us to considerably widen the range of values considered in the
$(L_\nu\langle E_\nu^2\rangle)$-$(\dot{M}M_\mathrm{NS})^{3/5}$-plane. The black-hole forming 2D and 3D models
partially exhibit a shock-expansion phase in which $\tau_\mathrm{adv}/\tau_\mathrm{heat}$ can exceed unity. This is caused
by violent SASI spiral and sloshing activity shortly before the neutron star becomes gravitationally unstable and its collapse
sets in. Despite therefore fulfilling the critical condition for shock expansion, we do not expect these models to produce 
supernova explosions. 

\begin{figure*}
\centering
 \includegraphics[width=0.63\textwidth]{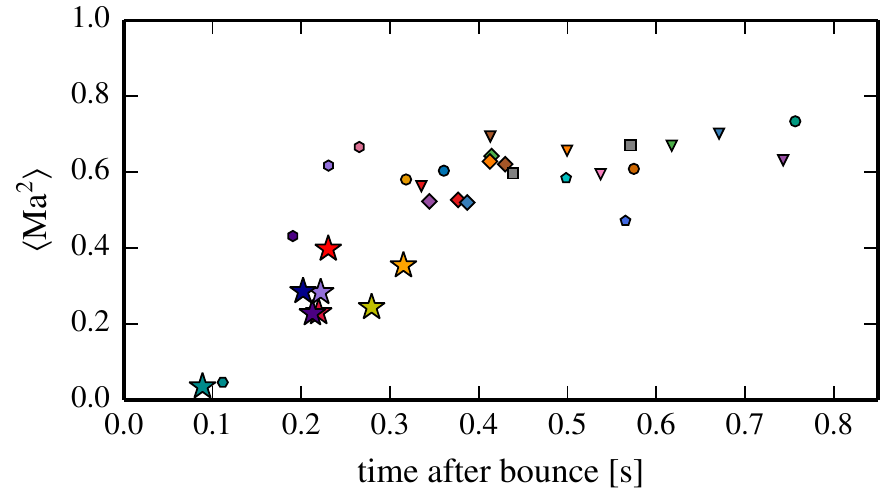}
 \caption{Average squared turbulent Mach number in the gain layer at the time 
  when the ratio $\tau_\mathrm{adv}/\tau_\mathrm{heat}$ reaches a value of 1.2 
  for all models displayed in Fig.\,\ref{crit_lum}. For the symbols and color coding,
  see caption of Fig.\,\ref{crit_lum}.\label{Mach}}
\end{figure*}

In the critical condition of Eq.\,(\ref{eq:nu_crit}), the critical value of the heating functional varies not only as a function
of $(\dot{M}M_\mathrm{NS})$, but also with the diverse effects accounted for by the factors assembled in $\xi_\mathrm{g}$ (Eq.\,\ref{eq:corr}),
which differ between progenitors and depend on the strength of nonradial instabilities, rotation, and on dimensionality.
Following \citet{Summa2016} and \citet{Janka2016}, we therefore construct a universal relation by normalizing 
$(L_\nu\langle E_\nu^2\rangle)$ by the factor $\xi_\mathrm{g}/\xi_\mathrm{g}^*$, where
\begin{equation}
\xi_\mathrm{g}^*\equiv \left.\left|\bar{e}_\mathrm{tot,g}\right|^{3/5}R_\mathrm{g}^{-2/5}\xi_\mathrm{turb}^{-3/5} \xi_\mathrm{rot}^{6/5}\right|^{\mathrm{s20}^\mathrm{LS}_\mathrm{strange}}_{\tau_\mathrm{adv}/\tau_\mathrm{heat}=1.2}
\end{equation}
is a constant introduced in order to define a reference case by the critical point of a (arbitrarily) chosen reference model.
Our universal critical relation therefore becomes
\begin{equation}
\left(L_\nu \langle E_\nu^2\rangle\right)_\mathrm{crit,corr}\equiv\frac{1}{\xi_\mathrm{g}/\xi_\mathrm{g}^*}\left(L_\nu \langle E_\nu^2\rangle\right)_\mathrm{crit}\propto(\dot{M}M_\mathrm{NS})^{3/5}.
\label{eq:crit_cond}
\end{equation}
for the ``corrected'' heating functional $(L_\nu \langle E_\nu^2\rangle)_\mathrm{crit,corr}$.

Note that, in contrast to \citet{Summa2016} and \citet{Janka2016}, we evaluate $\xi_\mathrm{g}^*$ here at the time when
$\tau_\mathrm{adv}/\tau_\mathrm{heat}=1.2$ (instead of unity). Considering our larger set of simulations it turns out that
some of them exceed a time-scale ratio of $\tau_\mathrm{adv}/\tau_\mathrm{heat}=1$ transiently although they do not reach explosion
conditions immediately. With the empirically chosen value of 1.2, we can well capture the critical behavior of all models 
of our set by the proportionality relation of Eq.\,(\ref{eq:crit_cond}) (see Fig.\,\ref{crit_lum}).
For the reference case, we picked the 3D model s20$^\mathrm{LS}_\mathrm{strange}$ \citep{Melson2015a}.
Besides the rotating models of the present work (m15\_3D\_artrot\_2deg as only exploding 3D case in red, m15\_3D\_rot\_2deg in brown, 
m15\_3D\_norot\_4deg in saddle brown; the exploding 2D models m15\_2D\_rot\_2deg and m15\_2D\_norot\_2deg are identified by gray squares
at their critical conditions), the list of 3D simulations in Fig.\,\ref{crit_lum} denotes models z9.6$^\mathrm{LS}$ and 
s20$^\mathrm{LS}_\mathrm{strange}$ as exploding cases of \citet{Melson2015,Melson2015a}, models s11$^\mathrm{LS}$, 
s20$^\mathrm{LS}$, and s27$^\mathrm{LS}$ as nonexploding simulations discussed by \citet{Tamborra2014a,Tamborra2014} and \citet{Hanke2013a},
models s9.0$^\mathrm{LS}_\mathrm{SMR}$ and s20$^\mathrm{LS}_\mathrm{strange,SMR}$ from \citet{Melson2016} and the black-hole
forming cases s40$^\mathrm{LS}$, u75$^\mathrm{LS}_\mathrm{SMR}$, u75$^\mathrm{LS}$, u75$^\mathrm{LS}_\mathrm{rep}$,
u75$^\mathrm{SFHo}$, and u75$^\mathrm{DD2}$ from Summa et al. (in preparation). 

Fig.\,\ref{crit_lum} shows the results of our analysis. In the upper panel, the uncorrected (open symbols) as well as the corrected values 
(filled symbols) for the heating functional at the onset of shock runaway are given for our set of exploding 3D (big asterisks) 
and 2D models (all other symbols). The correction procedure shows the desired effect: The scatter of the points is significantly
reduced (see especially the large corrections for the points in the right half of the plot and the bottom panel for a zoom 
to the lower left region of the top panel). 2D as well as 3D models can be fitted by a straight line which separates models 
with explosion or strong shock expansion from those with shock retraction.

When evolving along the evolution track (curves for $(L_\nu \langle E_\nu^2\rangle)_\mathrm{corr}$ versus $(\dot{M}M_\mathrm{NS})^{3/5}$
in the middle panel of Fig.\,\ref{crit_lum}; the evolution runs from the right to the left), the models hit the critical condition
for shock runaway at the moment they reach the critical line. The evolution tracks then (usually) bend sharply upwards because
the correction factor $\xi_\mathrm{g}/\xi_\mathrm{g}^*$ in the denominator on the rhs of Eq.\,(\ref{eq:crit_cond}) becomes very
small due to $\xi_\mathrm{turb}$ strongly increasing and $\left|\bar{e}_\mathrm{tot,g}\right|$ tending towards zero as more and more
matter in the gain layer becomes marginally bound \citep[for a detailed description, see][]{Summa2016}.

Our large sample of explosion models shows that the slopes of the critical lines for 2D (gray dashed line) and 3D (black dashed line) models are slightly different.
This is not surprising since Eq.\,(\ref{eq:crit_cond}) only provides a proportionality relation where the proportionality constant
is not fixed. Therefore differences in the explosion behavior of 2D and 3D simulations can manifest themselves in different slopes of
the best-fit straight lines. The separate fitting of 2D and 3D models also reduces the discrepancies between the universal critical curve
and the 3D data points that are visible at low $(L_\nu\langle E_\nu^2\rangle)_\mathrm{corr}$ and $(\dot{M}M_\mathrm{NS})^{3/5}$ values
in figure 3 of \citet{Janka2016}, which was only based on a reduced model set.

The difference of the critical curves for 2D and 3D models may be connected to a difference in the turbulent Mach number values at the beginning of explosion (see Fig.\,\ref{Mach}): 
In 2D, these values are (on average) a factor of about two higher than in 3D. This implies smaller values of $\xi_\mathrm{turb}$ (Eq.\,\ref{eq:xi_turb}) and therefore
bigger values of $\xi_\mathrm{g}$ (Eq.\,\ref{eq:corr}) for 3D models at their critical points, leading to smaller values of the corrected heating functional
$(L_\nu\langle E_\nu^2\rangle)_\mathrm{crit,corr}$ (Eq.\,\ref{eq:crit_cond}). In the black-hole forming massive stars this trend is
reversed, because in these cases extremely strong SASI activity and the associated kinetic energy pushes $\left|\bar{e}_\mathrm{tot,g}\right|$ closer to zero.
Our observation that 3D models can reach critical conditions for shock expansion in spite of smaller turbulent Mach numbers is compatible with 
similar findings by \citet{Mueller2016a} and \citet{Mueller2017}, who reported a smaller scaling factor between $E^\mathrm{aniso}_\mathrm{kin,g}/M_\mathrm{g}$
and $(\dot{Q}_\mathrm{heat}/M_\mathrm{g})^{2/3}$ in 3D compared to 2D. These facts could imply that in 3D other effects connected to 
turbulence might be more relevant than turbulent pressure, for example turbulent dissipation \citep{Mabanta2017} or more efficient convective heat
transport \citep[cf.][]{Murphy2013}. It is interesting that in the context of this discussion the ``two-dimensionalization'' of turbulent motions and 
of the corresponding power spectrum of the kinetic energy (Eq.\,\ref{eq:sph_har}, Fig.\,\ref{turb}) that is associated with
enhanced rotation in the case of model m15u6\_3D\_artrot\_2deg seems to manifest itself: In terms of turbulent Mach number, this model
(red asterisk in Fig.\ref{Mach}) is closest to the cloud of points of the 2D models, and also in Fig.\,\ref{crit_lum} the fast rotating 3D model
is very close to the critical line for the 2D cases.	

In view of the simple correction factors we applied to account for model-to-model variations with respect to nonradial mass motions, rotation, and further gain-layer
related quantities, the resulting critical lines capture the runaway behavior of all models remarkably well and confirm the existence
of a (slightly dimension-dependent) universal criterion for the development of critical conditions in our large set of 2D and 3D simulations.

\section{Conclusions} \label{sec:con}

Making use of three-flavor, energy dependent, ray-by-ray-plus neutrino
transport including the full set of state-of-the-art neutrino reactions and
microphysics,
we have studied the influence of rotation on the explosion dynamics of a 15\,M$_\odot$
progenitor \citep{Heger2005}
by hydrodynamic simulations of the neutrino-heating mechanism in 2D and 3D. 
Besides using the rotation profile provided by the progenitor model, in which
magnetic field effects on the angular momentum evolution had been taken into account
up to the onset of core collapse, we also performed simulations without rotation and
with a roughly 300 times higher angular momentum 
($\sim$$2\times 10^{16}$\,cm$^2$\,s$^{-1}$, corresponding to a spin period of
$\sim$20\,s,
instead of $\sim$$6\times 10^{13}$\,cm$^2$\,s$^{-1}$ and $\sim$\,6000\,s)
at the explosion-relevant Si/Si-O composition interface prior to collapse.

Our results suggest that sufficiently rapid rotation (as in our model with artificially
increased spin rate) favors the growth of strong SASI spiral activity in the 
equatorial plane and that the appearance of such SASI modes has an important influence
on the explosion. The dependence of success or failure on the stellar spin rate
is largely different between 2D (where nonaxisymmetric instabilities are absent)
and 3D, and it is also different from results of simplified toy-model studies (using
neutrino light bulbs) where feedback effects of rotation on the neutrino emission
are not taken into account.

``Moderate'' rotation as present in the original progenitor model has no significant
influence on the development of the supernova blast. In 2D, we obtained explosions
for the nonrotating as well as moderately rotating cases \citep[for the latter
case roughly 150\,ms earlier, but this is basically within the range of 
stochastic fluctuations between different 2D simulations with slightly varied conditions, see][]{Summa2016}. 
The 2D run with artificially increased rotation, however,
did not explode until nearly 500\,ms after core bounce, because
centrifugal effects on the postshock accretion flow and the centrifugal deformation
of the neutrinospheric layer reduce the neutrino luminosities and mean energies and thus the neutrino heating
in the gain layer. 

In 3D, in contrast, neither the nonrotating nor the moderately rotating models
showed any tendency to explode until $\sim$400\,ms after bounce, in line with
the 2D-3D differences previously obtained by the Garching group for 11.2, 20, 
and 27\,M$_\odot$ progenitors simulated with the same microphysics and numerical 
setup (except for the use of a Yin-Yang grid and a smaller 1D core of only
1.6\,km radius in the present 3D runs). Different from the 2D behavior, however,
the 3D simulation with the artificially increased rotation rate developed an 
explosion on a relatively short time scale of roughly 200\,ms after bounce, when
the infalling Si/Si-O interface reaches the shock wave. This success is 
facilitated by the strong spiral SASI motions of the postshock layer, which
store gravitational binding energy of the accretion flow in kinetic energy of
rotation. The spiral SASI pushes the shock outwards and thus increases the
volume and mass of the gain layer. This allows for a higher neutrino
heating rate and heating efficiency despite the rotationally reduced neutrino 
luminosities and mean energies. Moreover, it also stores gravitational binding 
energy of the accretion flow in kinetic energy of rotation and thus raises the
total specific energy of the matter in the gain layer, which drives the conditions
closer to the threshold for an explosion. These findings are consistent with 
conclusions drawn by \citet{Fernandez2015}.

The spiral SASI in our most rapidly spinning model supports shock expansion
most strongly in the equatorial plane, i.e., it triggers an oblate deformation, which
is in distinct contrast to simulations of explosions of fast rotating stars in 2D,
where a prolate shape of the outgoing blast wave is obtained (see Fig.\,\ref{entropy_cuts}). 
This fundamental difference was already pointed out by \citet{Nakamura2014}.
Interestingly, in our 3D model with rapid rotation and a Rossby number of
$\mathrm{Ro}\lesssim 1$, we observe a ``two-dimensionalization'' of the power
spectrum of the turbulent kinetic energy, which resembles the $\sim-3$ power-law 
decline at intermediate wavelengths that is characteristic of the 2D case (Fig.\,\ref{turb}). This
reflects the fact that the presence of centrifugal and coriolis forces constrains
the degrees of freedom for the turbulent fluid motions, and, similar to the 2D 
case, considerably more kinetic energy is stored on the largest possible scales
encompassed by the SASI spiral structures. 

Although at later times in our fast-spinning model a corotation 
radius between the SASI-deformed shock and the proto-neutron star surface
can be identified, interior to which the matter orbits
with shorter periods than the characteristic spiral-SASI pattern of the shock, our 
model does not develop a tightly-wound, one-armed spiral pattern as observed in
considerably faster spinning environments by \citet{Takiwaki2016} and
\citet{Kazeroni2017}. The small number of 3D simulations that we could 
perform with our detailed neutrino physics, as well as limitations of the 
numerical resolution that is affordable in these simulations (which might affect
the growth of instabilities to an extent we currently cannot quantify) do not
permit us to add substantial insights to a better and more systematic 
understanding of the subtle
interplay between rotation, the spiral SASI mode, and the corotational
instability as suggested by the recent works of \citet{Takiwaki2016},
\citet{Blondin2017} and \citet{Kazeroni2017}. Further studies of the 
different regimes of these instabilities are certainly desirable. 

The high rotation rate adopted for our exploding 3D model leads to a final neutron
star spin period of less than $\sim$5\,ms if angular momentum conservation is assumed.
This is considerably faster than the initial spin periods estimated for radio 
pulsars associated with supernova remnants, which are in the ballpark of several
tens to several hundreds of milliseconds \citep{Popov2012}. Since efficient spin-down
mechanisms of new-born neutron stars are not known \citep[see][]{Heger2005, 
Ott2006,Kazeroni2017}, the large angular momentum 
artificially imposed on the progenitor core of our fast rotating model does 
not seem to be representative of the birth sites for the majority of observed
neutron stars.

With a growing set of fully self-consistent 3D core-collapse simulations at hand, 
including previously published Garching models, the discussed models with rotation
of the present work, and more still unpublished results of nonrotating low-mass
as well as black-hole forming progenitors (Melson et al., in preparation; Summa
et al., in preparation, respectively), we were able to extend our analysis of
a critical luminosity condition for 2D models in \citet{Summa2016} to the
full set of 2D and 3D calculations. We confirmed our previous finding that a 
simple power-law 
relation (critical curve) between the ``corrected'' critical heating functional, 
$(L_\nu\langle E_\nu^2\rangle)_\mathrm{crit,corr}$, and the product
of preshock mass-accretion rate and neutron star mass,
$(\dot{M}M_\mathrm{NS})$, as expressed by Eq.\,(\ref{eq:crit_cond}), separates  the
domain where the supernova shock evolves as stalled accretion shock from the
domain of runaway shock expansion. This critical curve, which forms a straight
line in the
$(L_\nu\langle
E_\nu^2\rangle)_\mathrm{corr}$-$(\dot{M}M_\mathrm{NS})^{3/5}$-plane,
captures, with amazingly little scatter, the critical points for the transition 
between both regimes of shock behavior for progenitors from the 9\,M$_\odot$
range up to 75\,M$_\odot$ (Fig.\,\ref{crit_lum})\footnote{Investigated progenitors with 
40 and 75\,M$_\odot$
do not develop explosions, but can exhibit large-amplitude shock expansion driven by 
violent SASI activity prior to black-hole formation. Details will be reported in a 
forthcoming publication by Summa et al.\ (in preparation).}.
The correction factor applied to the heating functional accounts for effects of
turbulent pressure, rotation, and progenitor-dependent variations of the gain
radius and total specific energy of the material in the gain layer.

We diagnose a slight difference between the critical curves of 2D and 3D 
models: the critical condition for 3D models is somewhat steeper in the
$(L_\nu\langle
E_\nu^2\rangle)_\mathrm{corr}$-$(\dot{M}M_\mathrm{NS})^{3/5}$-space.
On the one hand this is connected to the fact that the turbulent 
Mach numbers in the gain layer at the time of shock runaway are
typically a factor of roughly two lower for 3D models (Fig.\,\ref{Mach}).
This decreases their correction factors in Eq.\,(\ref{eq:crit_cond}) and shifts the 3D critical line below
the 2D critical line for lower-mass stars (corresponding to the low side of the
$(\dot{M}M_\mathrm{NS})^{3/5}$ scale). On the other hand, in higher-mass
(black-hole forming) progenitors powerful sloshing and spiral SASI mass motions
drive the shock expansion and raise the (negative) total specific energy in the 
gain layer closer to zero by contributing significant amounts of kinetic energy.
This increases their correction factors and lifts the critical line for 3D models
above the one of the 2D counterparts on the high side of the 
$(\dot{M}M_\mathrm{NS})^{3/5}$ space. The difference in the turbulent
Mach numbers between 2D and 3D at the onset of shock runaway indicates that 
turbulent effects and nonradial mass motions manifest themselves in their 
effects on the supernova shock dynamics in different ways in 2D and 3D
simulations, potentially pointing to a higher relevance of turbulent pressure 
effects in 2D, whereas in 3D turbulent heat transport and spiral SASI activity 
may play a more important role. Consistent with the ``two-dimensionalization'' of the
turbulent power spectrum, rapid rotation with its constraints on the (turbulent) mass
motions in the gain layer moves our corresponding 3D model closer to the critical behavior
of the subset of 2D cases.

\acknowledgments

We are grateful to T.\ Foglizzo and R.\ Kazeroni for helpful discussions and
thank A.\ D\"oring for the visualizations of Figs.\,\ref{m15_rot} and \ref{m15_artrot}.
The project was supported by the Deutsche Forschungsgemeinschaft through 
the Excellence Cluster Universe (EXC 153; http://www.universe-cluster.de/)
and by the European Research Council through grant ERC-AdG No.\ 341157-COCO2CASA.
We acknowledge computing time through the European PRACE initiative on 
SuperMUC (GCS@LRZ, Germany) and MareNostrum (BSC-CNS, Spain) and through
the Gauss Centre for Supercomputing (GCS)
on SuperMUC at the Leibniz Supercomputing Centre (LRZ). 
Postprocessing was done on Hydra of the Max Planck Computing and Data Facility
(MPCDF).

\software{\textsc{Prometheus-Vertex} \citep{Fryxell1989,Rampp2002b,Buras2006a},
NumPy and Scipy \citep{Oliphant2007}, IPython \citep{Perez2007}, Matplotlib \citep{Hunter2007}.}

\bibliographystyle{apj}
\bibliography{ccsne}

\end{document}